\documentclass{aastex62}

\received{October 19, 2018}
\revised{}
\accepted{December 5, 2018}
\submitjournal{ApJ}

\shorttitle{MHD-Monte Carlo with feedback}
\shortauthors{Grimaldo et al.}

\begin{document}

\title{Proton acceleration in colliding stellar wind binaries}

\correspondingauthor{Emanuele Grimaldo}
\email{emanuele.grimaldo@student.uibk.ac.at}

\author{Emanuele Grimaldo}
\affiliation{Institute for Theoretical Physics \\
Technikerstra{\ss}e 21A \\
A-6020 Innsbruck, Austria}

\author{Anita Reimer}
\affiliation{Institute for Theoretical Physics \\
Technikerstra{\ss}e 21A \\
A-6020 Innsbruck, Austria}

\author{Ralf Kissmann}
\affiliation{Institute for Astro- and Particle Physics \\
Technikerstra{\ss}e 25 \\
6020 Innsbruck, Austria}

\author{Felix Niederwanger}
\affiliation{Institute for Astro- and Particle Physics \\
Technikerstra{\ss}e 25 \\
6020 Innsbruck, Austria}

\author{Klaus Reitberger}
\affiliation{Institute for Astro- and Particle Physics \\
Technikerstra{\ss}e 25 \\
6020 Innsbruck, Austria}

\begin{abstract}

The interaction between the strong winds in stellar colliding-wind binary (CWB) systems produces two shock fronts, delimiting the wind collision region (WCR). There, particles are expected to be accelerated mainly via diffusive shock acceleration (DSA). We investigate the injection and the acceleration of protons in typical CWB systems by means of Monte Carlo simulations, with both a test-particle approach and a non-linear method modelling a shock locally modified by the backreaction of the accelerated protons. We use magnetohydrodynamic simulations to determine the background plasma in the WCR and its vicinity. This allows us to consider particle acceleration at both shocks, on either side of the WCR, with a realistic large-scale magnetic field. We highlight the possible effects of particle acceleration on the local shock profiles at the WCR. We include the effect of magnetic field amplification due to resonant streaming instability (RSI), and compare results without and with the backreaction of the accelerated protons. In the latter case we find a lower flux of the non-thermal proton population, and a considerable magnetic field amplification. This would significantly increase the synchrotron losses of relativistic electrons accelerated in CWB systems, lowering the maximal energy they can reach and strongly reducing the inverse Compton fluxes. As a result, $\gamma$-rays from CWBs would be predominantly due to the decay of neutral pions produced in nucleon-nucleon collisions. This might provide a way to explain why, in the vast majority of cases, CWB systems have not been identified as $\gamma$-ray sources, while they emit synchrotron radiation.

\end{abstract}


\section{Introduction} \label{sec:intro}

Collisionless shocks are known to be sites where charged particles can be accelerated to relativistic energies. A fraction of the thermal particles of the plasma can be scattered by magnetic fluctuations and, after multiple crossings of the shock, gain energy by means of the first-order Fermi acceleration, in this context also known as diffusive shock acceleration (DSA) (e.g. \cite{drury1983, jones1991}).\\
An intriguing class of astrophysical objects, where particles are expected to be accelerated via DSA, are colliding-wind binaries (CWBs). These binary systems consist of hot, massive stars ejecting supersonic stellar winds, which eventually collide and form a (bow-shaped) wind-collision region (WCR), delimited by two shock fronts. Indeed, many CWBs have been identified as particle accelerators, mainly owing to the detection of a non-thermal radio component in their observed spectra (see e.g. \cite{debecker2013} and references therein). Up to now, only two such systems have been identified as $\gamma$-ray sources, namely $\eta$ Carinae \citep{reitberger2012,werner2013, reitberger2015} and $\gamma^2$ Velorum (also known as WR 11, \citet{pshirkov2016}). This is somewhat unexpected: several models (analytical and semi-analytical) predicted the production of $\gamma$-rays with fluxes above the detection threshold of \textit{Fermi}-LAT, HESS, MAGIC and VERITAS, mainly via inverse Compton scattering of electrons in the radiation field of the stars or due to the decay of neutral pions produced in hadronic interactions (e.g. \cite{eichler1993,benaglia2003,reimer2006}). These models have been improved recently by means of hydrodynamic (HD) and magnetohydrodynamic (MHD) simulations, combined with the solution of the transport equations for electrons and protons \citep{reitberger2014a,reitberger2014b,reitberger2017}. Such models allow to study these systems and possible effects of their geometry on the emission of $\gamma$-rays in a more detailed and realistic manner. A limitation of the method is the need to set an ``injection parameter'', which determines the fraction of thermal particles which can enter the acceleration process. Aiming at sidestepping this limitation, \cite{grimaldo2017a} combined MHD simulations of an archetypal CWB system, and Monte Carlo (MC) simulations of shock acceleration (similar to the one developed by \cite{ellison1995}). They found a high acceleration efficiency at the shocks of CWBs, suggesting that the backreaction of the accelerated particles on the shock structure must be taken into account.\\
A considerable amount of studies was conducted, focussing on the non-linear effects of particle acceleration at collisionless shocks, and several methods have been developed, mainly in the context of supernova remnants. For example, the Monte Carlo approach conserving momenta and energy fluxes was developed, amongst others, by \cite{ellison1984,vladimirov2009,bykov2014}. Semi-analytical methods \citep{amato2005,caprioli2009}, and time-dependent DSA simulations (e.g. \cite{kang2012}) have also been employed. A comparison between different methods for modelling shocks with non-linear DSA can be found in \cite{caprioli2010}. The most realistic approach for studying the injection mechanism is certainly the particle-in-cell (PIC) method (e.g. \cite{caprioli2014a}), which simulates the shock dynamics from first principles. However, it requires high computational costs. We therefore employ the MC technique to determine the fraction of accelerated particles with respect to the thermal plasma density. We use this injection parameter in the semi-analytical method developed by \cite{caprioli2009}, which includes the effects of the non-linear resonant instability. The equations of \cite{caprioli2009} are generalized for the case of oblique shocks. This is necessary for considering the variety of shock obliquities along the WCR.\\
In the next two sections, we will introduce the most important equations describing the numerical method. In Section \ref{sec:validation} we will compare the results for strictly parallel shocks to those for oblique shocks, with parameters typical of SNRs (used as a reference). In Section \ref{sec:results} we will apply the non-linear MHD-MC method to an archetypal CWB system, considering different positions along both shocks delimiting the WCR and comparing test-particle to non-linear results. The last section is devoted to the conclusions.

\section{Theoretical background} \label{sec:theory}

In order to ensure a better comprehension of the problem, we review the most important equations and definitions. Different methods employed for modelling non-linear shocks show some common features concerning the modification of the velocity profile: a shock precursor develops, where the inflowing plasma is progressively slowed down by the pressure of the non-thermal protons, up to the position of the MHD shock, usually called ``subshock''. The magnetic turbulence generated by the backstreaming charged particles during the acceleration process can have an effect on the shock structure \citep{caprioli2009}. Therefore, we will use the MHD equations including the turbulence pressure. For the sake of simplicity, we consider only Alfv\'en waves generated by the resonant streaming instability. We use here the indices 0, 1 and 2 for quantities far upstream, directly upstream and downstream of the subshock, respectively. Let us suppose that the only spatial dependence of the background fields (flow velocity $\bm{u}$, magnetic field $\bm{B}$, density $\rho$, and temperature $T$) is in the $x$-direction. Following \cite{scholer1971} and \cite{decker1988}, we write the MHD conservation laws in the presence of Alf\'en waves as:
\begin{subequations}
\begin{align}
		[\rho u_x]_1^2&=0 \ , \\
		\left[\rho \bm u u_x + \left(p_g + \frac{B^2}{2\mu_0} + P_w \right) \hat{\bm{n}} - \frac{B_x \bm{B}}{\mu_0} \right]_1^2&=0 \ , \\
		\left[u_x \left\{ \frac{1}{2} \rho u^2 + \frac{\gamma}{\gamma-1} p_g + \frac{B^2}{\mu_0} + F_w\right\}-\frac{B_x(\bm{B}\cdot\bm{u})}{\mu_0} \right]_1^2&=0 \ , \label{eq:MHDEFluxConservation}\\
		[B_x]_1^2&=0 \ , \\
		[\hat{\bm{n}}\times(\bm{u}\times\bm{B})]_1^2&=0 \ .
\end{align}
\label{eq:MHDConservation}
\end{subequations}
\noindent Herein, $\hat{\bm{n}}$ is the unit vector normal to the shock, $\mu_0$ is the permeability constant, $\gamma$ is the adiabatic index, $p_g=nk_BT$ is the thermal pressure of the plasma, where $T$ is its temperature, $n$ is the particle density, and $k_B$ is the Boltzmann constant. A subscript $x$, $y$ or $z$ indicates the $x$-, $y$- or $z$- component of a vector, while the notation $[]^2_1$ denotes the difference between downstream and upstream quantities. The pressure term associated with the Alfv\'en waves is $P_w=(\delta B)^2/(2\mu_0)$, where $\delta B$ is the magnetic field amplitude of the wave. The energy flux $F_w$ in Eq. \eqref{eq:MHDEFluxConservation} includes the kinetic energy flux and the $x$-component of the Poynting vector associated to the wave:
\begin{equation}
F_w=\frac{1}{2} \rho (\delta u)^2 u_x + \frac{1}{\mu_0} \{ (\bm{B}\times\bm{\delta u} + \bm{\delta B}\times\bm{u})\times\bm{\delta B} \}\cdot \hat{\bm{n}}  \ ,
\label{eq:Fw}
\end{equation}
where $\bm{\delta u}$ is the velocity change of the plasma due to the Alfv\'en waves.
By using the transmission and reflection coefficients of Alfv\'en waves incident on a shock given by \cite{mckenzie1969}, \cite{scholer1971} obtain a third-order equation for the compression ratio $r\equiv\rho_2/\rho_1=u_{1x}/u_{2x}$ equivalent to:
\begin{equation}
a_3 r^3 + a_2 r^2 + a_1 r + a_0 =0\ ,
\label{eq:thirdr}
\end{equation}
with coefficients
\begin{eqnarray*}
a_3 &=& [(\gamma-1)(1+\lambda) M_{A1x}^2 + \gamma \beta_1 \cos^2\theta_{B1} ] \cos^2\theta_{B1}\\
a_2 &=& \{ [2(1+\lambda)-\gamma (1+\cos^2\theta_{B1}+\lambda)] M_{A1x}^2 - [1+\lambda+\gamma(2\beta_1+1+\lambda)]\cos^2\theta_{B1} \} M_{A1x}^2\\
a_1 &=& [(\gamma-1)M_{A1x}^2 + \gamma (1+\lambda+\cos^2\theta_{B1}+\beta_1) + 2 \cos^2\theta_{B1}] M_{A1x}^4\\
a_0 &=& -(\gamma+1) M_{A1x}^6
\end{eqnarray*}
Herein,  $\theta_B$ is the angle between the shock normal and the magnetic field, $\lambda=(\delta B_1 / B_1)^2$, $M_{A1x}=u_{1x}/v_A$, with the Alfv\'en speed $v_A=B/\sqrt{\mu_0\rho}$, and $\beta=p_g 2\mu_0/B^2$. In the limit $\lambda=0$ Eq. \eqref{eq:thirdr} reduces to Eqs. (11) of \cite{decker1988}.\footnote{This is true after correcting for a typo in \cite{decker1988}: the term $\cos^2\delta_1$ should always be multiplied by $M_{A1}^2$ in their equations (11a)-(11c).} Using Eqs. \eqref{eq:MHDConservation}, one can find the expressions relating the variations of the magnetic field, the flow velocity and the density along $x$, assuming that the magnetic field lies in the $x$-$z$ plane:
\begin{subequations}
\begin{align}
u_x(x)&=\frac{\rho_0 u_{0x}}{\rho(x)}\ , \label{eq:ux(x)} \\
u_y(x)&=u_{0y}\ , \label{eq:uy(x)}\\
u_z(x)&=u_{0z}+\left( \frac{B_z(x)-B_{0z}}{\mu_0} \right) \frac{B_{0x}}{\rho_0 u_{0x}} \ ,
\label{eq:uz(x)}
\end{align}
\label{eq:u(x)}
\end{subequations}
\begin{subequations}
\begin{align}
B_x(x)&=B_{0x}\ , \label{bx(x)}\\
B_z(x)&=\left (  \frac{M_{A0x}^2-\cos^2\theta_{B0}}{U_x(x) M_{A0x}^2 - \cos^2\theta_{B0}} \right )\ B_{0z} \ .
\label{eq:Bz(x)}
\end{align}
\label{eq:B(x)}
\end{subequations}
\noindent Here and in the following sections, velocities and pressures indicated with capital letters are normalized by $u_{0x}$ and $\rho_0 u_{0x}^2$, respectively. The electric field is $\bm{E}=-\bm{u}\times\bm{B}$, and it is entirely due to the motion of the plasma in the magnetic field. In all of our set-ups, $\bm{u}$ is (almost) perfectly parallel to $\bm{B}$, and the electric field is therefore small.\\
For strong turbulence, $\delta B/B>1$, the definition of a ``background magnetic field'' becomes questionable. In such cases, we consider $\bm{B}(x)$ just as the field determining the direction of propagation of the Alfv{\'e}n waves.

\section{Numerical Methods} \label{sec:methods}

Our method combines MHD simulations of the wind plasma, obtained with the \textsc{Cronos} code \citep{kissmann2018, kissmann2016}, Monte Carlo simulations of shock acceleration, using the technique developed by e.g. \cite{ellison1995}, and a semi-analytical method for computing the local non-linearly modified shock profile \citep{amato2005}. These components can be run independently from each other. As far as the MHD code is concerned, we refer the reader to the above-cited references. Below we provide details concerning the Monte Carlo part, the semi-analytical part, and how the different components are combined.
\subsection{Monte Carlo simulations}\label{subsec:MC}

The Monte Carlo method is similar to that of \cite{ellison1995}. We let each particle move using the Bulirsch-Stoer algorithm \citep{press1992}. The particles are acted upon by the Lorentz force given by the background electromagnetic fields. After a time $t_c$, exponentially distributed with a mean value \mbox{$\bar{t}_c=\xi r_g/v$}, a scattering occurs (elastic in the frame of the local plasma flow). Here, $r_g=p/(qB)$ is the gyroradius, where $p$ is the magnitude of the momentum of a particle, $q$ is its charge, $v$ is its speed, $B$ is the local magnetic field strength, and $\xi$ is a proportionality factor relating gyroradius and mean free path ($\lambda_\textrm{mfp}=\xi r_g$). In the following, we set $\xi=1$, corresponding to Bohm diffusion. This choice is supported by PIC simulations \citep{caprioli2014c}. The new direction of the momentum vector is randomly determined at each scattering, mimicking strong magnetic turbulence.\\
The protons of the background plasma are assumed to have a Maxwell-Boltzmann distribution in the local plasma frame, and are injected accordingly, close to the shock front, following the prescription of \cite{vladimirov2009}. The densities $n$ and fluxes $\Phi$ at the shock fronts are given by:
\begin{equation}
	\begin{aligned}
		n&=\sum_i \left | \frac{u_{0}}{v_{x,i}}\right | w \\
		\Phi&=\sum_i v_{x,i} \left | \frac{u_{0}}{v_{x,i}}\right | w \ .
	\end{aligned}
\label{eq:flux_and_density}
\end{equation}
Herein, $u_{0}$ is the flow speed, $w=n_{0}/N_p$, where $n_{0}$ is the particle density and $N_p$ is the number of particles injected, and $v_{x,i}$ is the $x$-component of the velocity of the particle crossing the measurement surface (i.e. the shock). The quantities $u_0$ and $w$ refer to the point where the particles are injected at the beginning of the simulation. The index $i$ runs over all the crossing events of all the simulated particles. As discussed by \cite{vladimirov2009} (see Eq. (3.9) of that work), if $\Delta p_k$ is the width of the $k$-th momentum bin centred at momentum $p_k$, the distribution function for particles passing a surface of interest is:
\begin{equation}
		f(p_k)=\frac{1}{4\pi p_k^2 \Delta p_k}\ n(p_k) \ ,
\label{eq:distribfunciton}
\end{equation}
with
\begin{equation}
	n(p_k) = \sum_{p_i\in\Delta p_k} \left | \frac{u_{0}}{v_{x,i}}\right | w \ .
\label{eq:defn1}
\end{equation}
Here, the index $i$ runs over all the crossing events of the particles having a momentum within the $k$-th momentum bin.

In order to use the MHD results as the background for the Monte Carlo simulations, we first locate the position of the shocks by setting a threshold for the gradient of the temperature, which abruptly rises there from $\sim 10^4$ K to $10^7 - 10^8$ K. A system of superimposed cells (super-cells) is then initialized, with the purpose of (i) having a sharp shock jump for thermal particles, and (ii) avoiding artefacts in the acceleration of particles due to misalignment of the shock surface and the boundary between upstream and downstream cells. In fact, the simulated protons would not ``see'' a sharp shock, because the size of the MHD cells is much larger than the mean free path of the thermal particles, and the transition from upstream to downstream is about three cells wide.\footnote{In this work, the MHD cells are cubes of edge length $\Delta x=3.9\mbox{ R}_{\sun}$.} A more detailed description of the procedure can be found in Appendix \ref{sec:app_bg_treatment}.\\
We inject the particles close to the shock in one selected upstream super-cell and let them move and scatter from then onwards. We stop the simulation of a particle when (i) it reaches a distance $x_D=10 D/u_2$ downstream of the shock, where $D$ is the diffusion coefficient given by Eq. \eqref{eq:diffOblique}, when still in the initial super-cell system, or (ii) it leaves the whole simulated box, or (iii) the number of scatterings experienced by the particle reaches a pre-set value, much greater than the expected mean number of scatterings needed to reach the highest possible energy in the system. Assuming an infinitely extended shock downstream, as it is effectively for low-energy (thermal) particles, the choice $x_D=10 D/u_2$ corresponds to stopping the simulation for a particle when its probability to return to the shock is $\lesssim e^{-10}$\citep{ostrowski1993}.\\
In the next section, we will describe the semi-analytical method used for the determination of the local non-linear modifications of the shock.

\subsection{Semi-analytical non-linear calculations}\label{subsec:semi_anNL}

\begin{figure}
\centering
\includegraphics[width=.33\columnwidth]{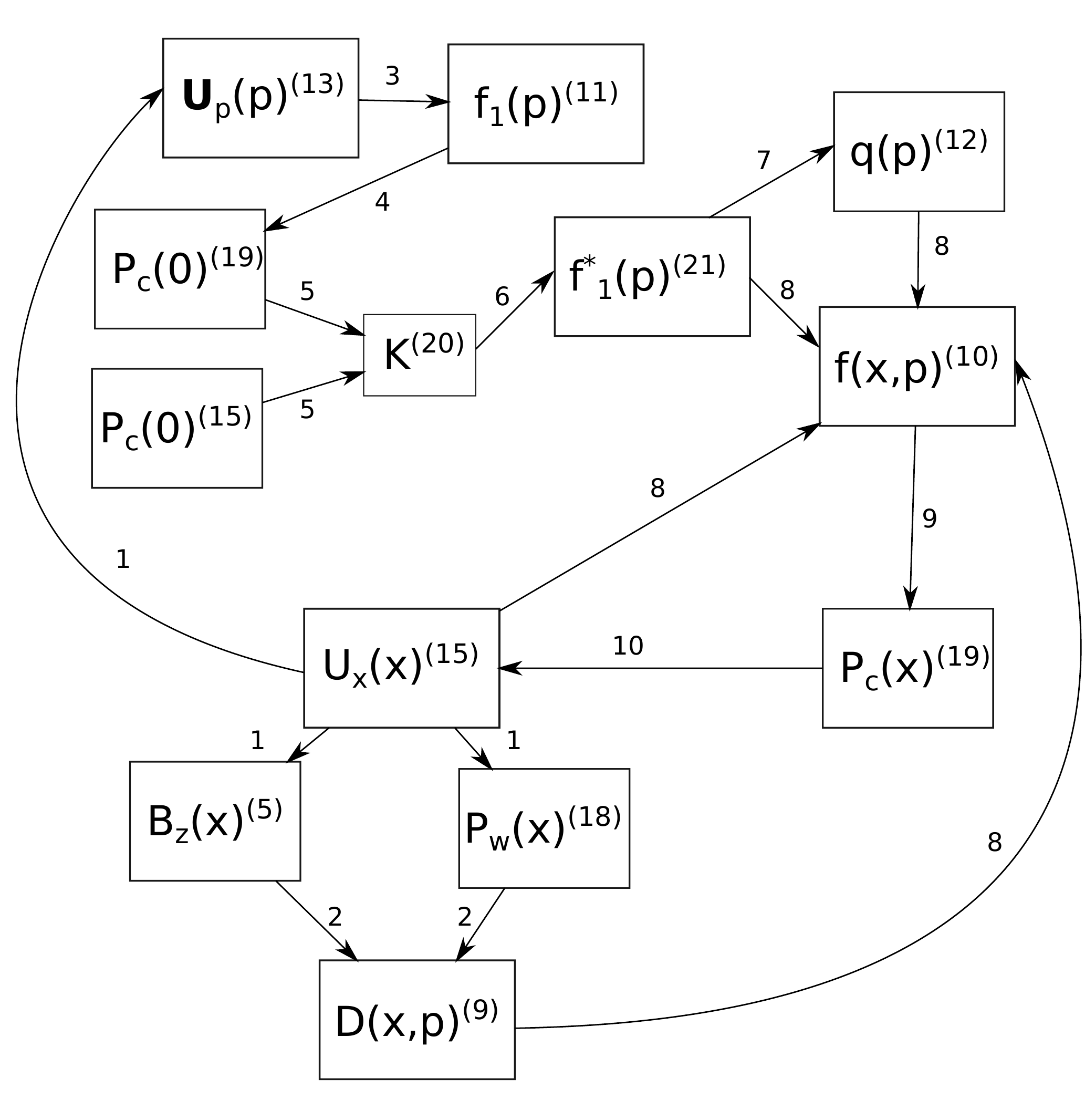}
\caption{Scheme of the algorithm used for the calculation of the non-linearly modified shock profile at a fixed compression ratio $r_\textrm{sub}$. The indices at the exponent indicate the equation needed. The numbers of the arrows indicate the step within the single iteration.}
\label{fig:algorithm}
\end{figure}

\begin{figure}[ht!]
\centering
\includegraphics[width=.43\columnwidth]{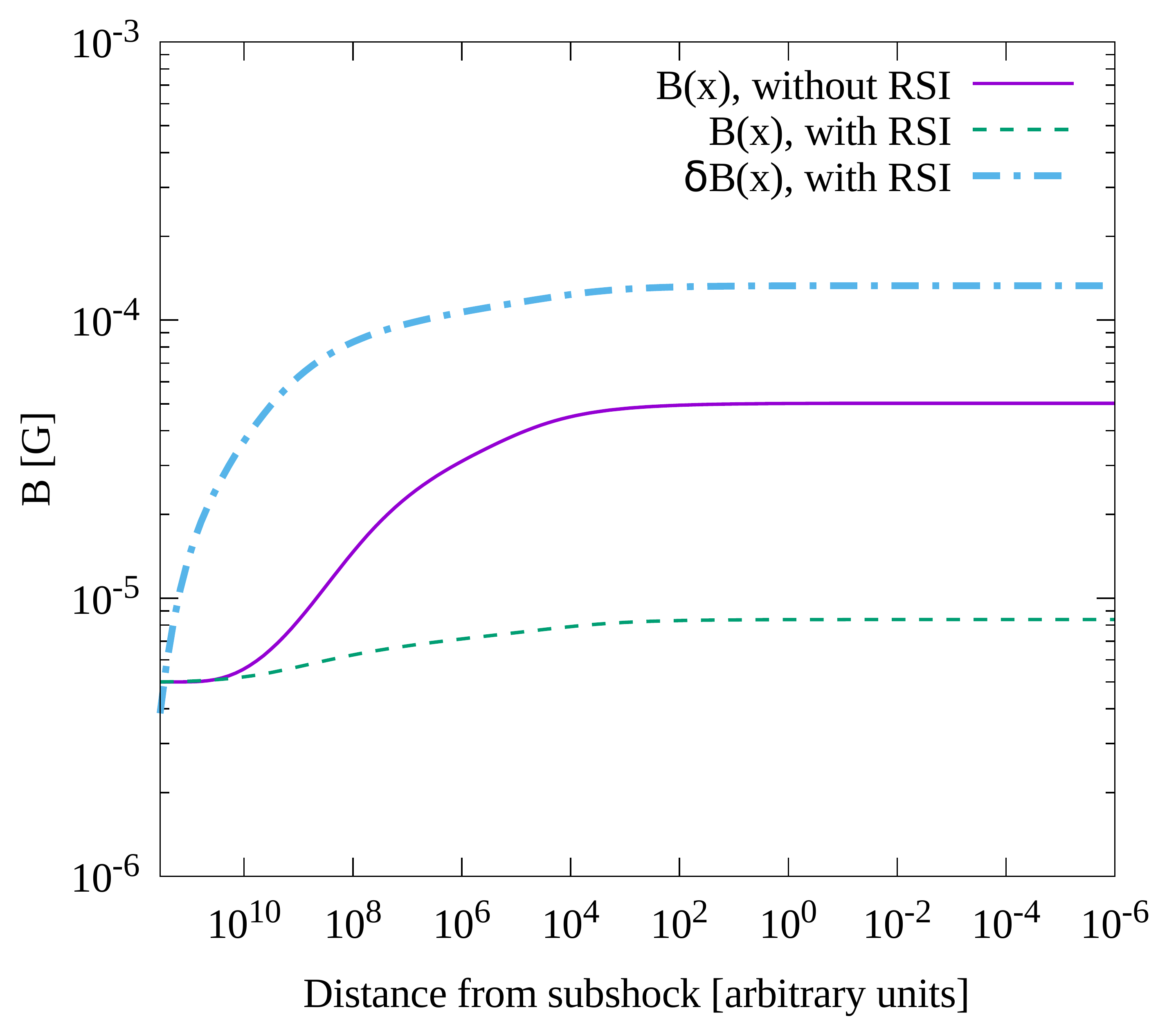}
\caption{Magnitudes of the background and RSI magnetic fields upstream, as a function of the distance from the subshock, for shock obliquity $\theta_{B0}=30^\circ$. $B(x)$ is the absolute value of the background magnetic field. In the case without RSI (solid line), its increase is due to the increase of $B_z$, caused by the decreasing plasma speed (see Eq. \eqref{eq:Bz(x)}). The magnetic field $\delta B(x)$ associated with the Alfv\'en waves generated via RSI (dot-dashed line) strongly reduces the compression ratio, and therefore the ratio between the far upstream plasma speed and that at the subshock. Therefore, the increase of $B(x)$ (dashed line) is much smaller when RSI is taken into account.  \label{fig:Bvsx}}
\end{figure}

In order to obtain a shock profile which conserves energy and momentum fluxes, we adapt the procedure developed by \cite{caprioli2009} to the case of oblique shocks. It consists of an iterative method solving the diffusion-advection equation for the accelerated particles at the shock. The background conditions are determined by the velocity $\bm u$, the density $\rho$, the temperature $T$, and the magnetic field $\bm B$. We assume that all the considered quantities change locally only in the $x$-direction, and that $v_A\ll u$. We can then write:
\begin{equation}
u_x(x) \frac{\partial f(x,p)}{\partial x} = \frac{\partial}{\partial x} \left [ D(x,p) \frac{\partial f(x,p)}{\partial x} \right ] + \frac{\textrm{d}u_x(x)}{\textrm{d}x} \frac{p}{3} \frac{\partial f(x,p)}{\partial p} + Q(x,p)\ ,
\label{eq:diff-adv}
\end{equation}
where $f(x,p)$ is the isotropic part of the distribution function of the accelerated particles. The source term, which accounts for the injection of particles in the acceleration process, is given by:
\begin{equation}
Q(x,p)=\frac{\eta \rho_1 u_1}{4\pi m_p p^2_\textrm{inj}} \delta(p-p_\textrm{inj}) \delta(x)\ ,
\label{eq:sourceTerm}
\end{equation}
where $m_p$ is the proton mass, the $\delta$ is the Dirac delta distribution, $p_\textrm{inj}$ is the ``injection momentum'', and $\eta$ is the ``injection efficiency'' (see below for more details concerning these last two terms).
The diffusion coefficient is (e.g. \cite{jones1991}):
\begin{equation}
	\begin{aligned}
		D(x,p)&=D_\parallel(x,p) \cos^2\theta_B(x) + D_\perp(x,p) \sin^2\theta_B(x)\ , \\
		D_\parallel(x,p)&=\xi \frac{p}{qB(x)} \frac{v}{3} \ , \\
		D_\perp(x,p)&=\frac{\xi}{(1+\xi)^2} \frac{p}{qB(x)} \frac{v}{3} \ ,
	\end{aligned}
\label{eq:diffOblique}
\end{equation}
$D_\parallel$ and $D_\perp$  being the diffusion coefficients parallel and perpendicular to the magnetic field lines, respectively.
The solution to Eq. \eqref{eq:diff-adv} is given by (see, e.g., \cite{amato2005}):
\begin{equation}
f(x,p)=f_1(p)\exp\left\{-\frac{q(p)}{3} \int_{-\infty}^0{\textrm{d}x' \frac{u_x(x')}{D(x',p)} }\right\} \ ,
\label{eq:fxp}
\end{equation} 
where
\begin{equation}
f_1(p) = \frac{\eta n_0}{4\pi p_\textrm{inj}^3} \frac{3\ r_\textrm{tot}}{r_\textrm{tot} U_{px}(p)-1} \exp\left [-\int_{p_\textrm{inj}}^p \frac{dp'}{p'} \frac{3 r_\textrm{tot} U_{px}(p')}{r_\textrm{tot} U_{px}(p')-1} \right ] \ 
\label{eq:f1}
\end{equation}
is the distribution function immediately upstream of the subshock, with 
\begin{equation}
q(p)=-d \log f_1(p)/ d\log p\ .
\label{eq:qp}
\end{equation} 
Here, $r_\textrm{tot}=u_{0x}/u_{2x}$ is the total compression ratio of the shock, while the mean velocity of the scattering centres (of the plasma) ``seen'' by a particle of momentum $p$ is:
\begin{equation}
\bm{u_p}(p)=\bm{u_{1}}-\frac{1}{f_1(p)}\int_{-\infty}^0{\textrm{d}x \frac{\textrm{d}\bm u(x)}{\textrm{d}x} f(x,p)} \ .
\label{eq:up}
\end{equation}
Similarly, for the magnetic field we have:
\begin{equation}
\bm{B_p}(p)=\bm{B_{1}}-\frac{1}{f_1(p)}\int_{-\infty}^0{\textrm{d}x \frac{\textrm{d}\bm B(x)}{\textrm{d}x} f(x,p)} \ .
\label{eq:Bp}
\end{equation}
Accordingly, the ``mean'' electric field is: $\bm{E_p}=-\bm{u_p}(p)\times\bm{B_p}(p)$.
The momentum flux conservation equation, normalized by the kinetic momentum flux, reads:
\begin{equation}
1 + P_{g0} + P_{B0} = U_x(x) + P_g(x) + P_w(x) + P_B(x) + P_c(x) \ .
\label{eq:momFluxConsNorm}
\end{equation}
Here, $P_g$ is the thermal pressure of the plasma, $P_w$ is the pressure associated with the Alfv{\'e}n waves produced by the resonant streaming instability, $P_B$ is the background magnetic field pressure due to the $z$-component of the field, and $P_c$ is the pressure of the accelerated protons. Considering only adiabatic heating in the precursor, one has:
\begin{equation}
P_g(x)=\frac{U_x(x)^{-\gamma}}{\gamma M_{0x}^2} \ ,
\label{eq:Pg}
\end{equation}
where $M_{0x}^2=\rho_0 u_{0x}^2/(\gamma p_{g0})$.
The $z$-component of the magnetic field exerts a pressure:
\begin{equation}
P_B(x)=\frac{B_z^2(x)}{2\mu_0 \rho_0 u_{0x}^2} \ ,
\label{eq:PB}
\end{equation}
where the $B_z$ can be found using Eq. \eqref{eq:Bz(x)}. This pressure term is present only if the shock is not strictly parallel, and is usually negligible in the precursor. However, as we will see, it has an influence on the jump conditions at the subshock and on the total compression ratio in the case of efficient particle acceleration, especially in the absence of magnetic field amplification. 
The term $P_w$, the pressure due to Alfv{\'e}n waves, is given by (see Appendix \ref{sec:app_wavepressure} for a brief derivation):
\begin{equation}
P_w(x)=\frac{U_x(x)^{-\frac{3}{2}}}{4 M_{A0x}}\left [ (1-U_x^2(x))\ \cos\theta_{B0}  \right ] \ .
\label{eq:alpha(x)}
\end{equation}
In order to find the non-linearly modified shock profile, we proceed as illustrated by \cite{caprioli2009},
using the equations adapted for the case of oblique shocks. We first set the compression ratio at the subshock, $r_\textrm{sub}$, and therefore the total compression ratio $r_\textrm{tot}$ (see below for more details concerning this point). A scheme of the algorithm can be seen in Figure \ref{fig:algorithm}.
In the first iteration we set $u_{px}(p)=u_x(x)=u_{1x}$. Starting with a different value, e.g. $u_{0x}$ does not affect the final result. The magnetic field, the wave pressure and the diffusion coefficient are calculated according to Eqs. \eqref{eq:Bz(x)}, \eqref{eq:alpha(x)} and \eqref{eq:diffOblique}. We compute $f_1(p)$ using Eq. \eqref{eq:f1}, which in turn allows to compute $P_{c1}$, i.e. $P_c(x)$ at the subshock, according to:
\begin{equation}
P_c(x)= \frac{4\pi}{3\rho_0 u_0^2} \int_{p_\textrm{inj}}^{p_\textrm{max}} dp \ p^3 v(p) f(x,p) \ .
\label{eq:PcAlternative}
\end{equation}
At this point, we calculate the pressure of the accelerated particles also with Eq. \eqref{eq:momFluxConsNorm}, and we find 
\begin{equation}
K=P_{c1}^{\eqref{eq:momFluxConsNorm}}/P_{c1}^{\eqref{eq:PcAlternative}}\ , 
\label{eq:K}
\end{equation}
where the bracketed exponent indicates the equation used for the computation. This allows us to normalize $f_1(p)$ and obtain
\begin{equation}
f_1^*(p)=K\ f_1(p)\ , 
\label{eq:fstar}
\end{equation}
so that the momentum flux between far upstream and the subshock is conserved. This normalized distribution function is used to calculate $f(x,p)$ by means of Eq. \eqref{eq:fxp}, and $P_c$ by means of Eq. \eqref{eq:PcAlternative}. Finally, the velocity profile $U_x(x)$ can be obtained using Eq. \eqref{eq:momFluxConsNorm}, which allows to find $\bm B(x)$ and $P_w(x)$, and in turn the diffusion coefficient $D(x,p)$ using Eqs. \eqref{eq:Bz(x)}, \eqref{eq:alpha(x)} and \eqref{eq:diffOblique}, respectively. A new iteration is then started by calculating $u_{px}(p)$ by means of Eq. \eqref{eq:up}. In order to achieve faster convergence, we average the flow profile between iteration $n$ and $n-1$, before computing $\bm B(x)$, $D(x,p)$ and $u_{px}(p)$. A similar solution has also been used by \cite{amato2005} (private communication). We stop the cycle when $K$ does not change more than a specified amount between consecutive iterations. At this point, we check the value of $K$: if it is within a $15\%$ tolerance interval around 1, the solution has been found, otherwise a new $r_\textrm{sub}$ is used and the procedure is repeated (see Section \ref{sec:validation} for a discussion on the convergence criterion). The compression ratio is increased if $K>1$, while it is decreased if $K<1$.\\
 Eq. \eqref{eq:thirdr} allows us to compute the flow velocity and magnetic field downstream, when combined with Eqs. \eqref{eq:Bz(x)} and \eqref{eq:uz(x)}, once the conditions at the subshock are known. The temperature downstream is given by the relation:
\begin{equation}
\frac{T_2}{T_1} = \frac{1}{r} \frac{p_{g2}}{p_{g1}} = \frac{1}{r[(\gamma+1)-(\gamma-1)r]} \left \{ (\gamma+1)r - (\gamma-1) + (\gamma-1) \frac{M_{A1x}^4(r-1)^3}{(M_{A1x}^2-r\cos^2\theta_{B1})^2} \frac{(p_{B1}+p_{w1})}{p_{g1}} \right \} \ .
\label{eq:T2}
\end{equation}
This equation can be obtained with the same approach as used by \cite{vainio1999}, except using the conservation laws and transmission and reflection coefficients of Alfv\'en waves for the case of oblique shocks given by \cite{scholer1971}.\\
In order to find $r_\textrm{tot}$, we numerically solve Eq. \eqref{eq:thirdr} keeping $r=r_\textrm{sub}$ fixed and employing the relation $U_{1x}=r_\textrm{sub}/r_\textrm{tot}$. In this way, knowing the far upstream conditions and $r_\textrm{sub}$, we can determine the background directly upstream and downstream of the subshock. At this point, the only missing ingredient is the fraction $\eta$ of particles being injected into the acceleration process. \cite{caprioli2009} use the formula:
\begin{equation}
\eta=\frac{4}{3\sqrt{\pi}}(r_\textrm{sub}-1)\psi^3 e^{-\psi^2} \ .
\label{eq:etaCapr}
\end{equation}
They consider a Maxwell-Boltzmann distribution with the temperature of the shocked plasma and assume that only the particles with momentum $p_\textrm{inj}\geq \psi p_\textrm{th,2}$, with $p_\textrm{th,2}=\sqrt{2mk_BT_2}$, can be injected into the Fermi acceleration. This solution aims at modelling the thickness of the shock, assuming that particles with gyroradii smaller than the shock thickness can only be advected away from the shock and will not contribute to the non-thermal tail of the particle distribution function. Values of $\psi\simeq2-4$ are usually chosen. \cite{blasi2005} showed that, for example, $\psi\approx2$ corresponds to a shock thickness $\lambda_{sh}=r_g^\textrm{th,2}$, and $\psi\approx3.25$ corresponds to $\lambda_{sh}=2r_g^\textrm{th,2}$, where $r_g^\textrm{th,2}$ is the gyroradius of a particle with momentum $p_\textrm{th,2}$.\\
In this work, we choose a more accurate way to determine $\eta$ by means of Monte Carlo simulations. At the beginning of the first cycle associated to the first guess for the compression ratio $r_\textrm{sub}$, we roughly estimate the ``injection efficiency'' as follows. We initialize the background for MC simulations with the parameters of the subshock (i.e. $\bm u_1$, $\bm B_1$, etc., upstream and $\bm u_2$, $\bm B_2$, etc., downstream). We then inject the particles upstream, close to the shock, and determine $\eta$ as $\eta=n_\textrm{ret}/n_\textrm{tot}$, where $n_\textrm{ret}$ is the number of particles recrossing the shock from downstream to upstream, and $n_\textrm{tot}$ is the total number of injected particles. Using the injection efficiency so obtained, we calculate a new $\psi$ which satisfies Eq. \eqref{eq:etaCapr}, and a new $p_\textrm{inj}=\psi p_\textrm{th,2}$. 
We thus find the modified shock solution for the current $r_\textrm{sub}$ employing the semi-analytical method described above. Finally, in order to obtain a more accurate estimate for the injection efficiency, and in turn for the density of the non-thermal population, which is essential for obtaining an energy-conserving solution, we run Monte Carlo simulations letting particles reach momenta $p\approx100 p_\textrm{inj}$. The particle density obtained from the Monte Carlo runs is then compared, at the momentum $p=10 p_\textrm{inj}$, to the particle density obtained from the semi-analytical calculations, in order to find a corrected $\eta$ and the respective $K$. The comparison is done at $10 p_\textrm{inj}$ in order to avoid the low-energy part of the spectrum after the thermal peak, which shows some oscillations, and the cut-off at the end of the distribution. A more accurate estimation would require to simulate the spectra up to higher energies, since there can be a difference in the slopes, up to momenta of $p\approx m_p c$, of the non-thermal distributions obtained with the Monte Carlo technique as compared to the ones obtained with the semi-analytical method. This discrepancy, ascribed to different treatment of the transition between thermal and non-thermal particles, was also found in \cite{caprioli2010}. It has been shown in the same work that the spectra at high energies (i.e. above momenta $p\approx m_p c$) are in good agreement. For the purposes of this paper, the possible gain in accuracy in the determination of the normalization of the non-thermal distributions does not justify the related increase of the computation times.\\
For technical reasons, instead of subdividing the computational box upstream into many cells, we use only two cells (upstream and downstream). The shock modification is taken into account by using the momentum-dependent averaged quantities $\bm{u_p}(p)$, $\bm{B_p}(p)$: when a particle of momentum $p$ is in the upstream cell, we use the background fields $\bm{u_p}(p)$, $\bm{B_p}(p)$, given by Eqs. \eqref{eq:up} and \eqref{eq:Bp}, and the corresponding $\bm{E_p}(p)$. \cite{caprioli2009} chose to use the amplified magnetic field for computing the diffusion coefficient, i.e. they apply Eq. \eqref{eq:diffOblique} using $B(x)=\delta B(x)\equiv \sqrt{2\mu_0 P_w(x) \rho_0 u_0^2}$. In Figure \ref{fig:Bvsx} we show an example of how the magnetic field varies with the distance from the shock. Far upstream, the particle density is $n_0=0.5\times10^6$ m$^{-3}$, the magnitude of the magnetic field is $B_0=5\times 10^{-11}$ T, the temperature is $T_0=10^4$ K, and the plasma speed is $u_0=5.9\times10^6$ m s$^{-1}$. We show the case with shock obliquity $\theta_{B0}=30^\circ$ (see also Table \ref{tab:comparison}). The change of $B(x)$ is due to the change in the $z$-component of the magnetic field (Eq. \eqref{eq:Bz(x)}), while $\delta B(x)$ changes according to Eq. \eqref{eq:alpha(x)}. The diffusion coefficient is therefore clearly changing with the distance from the subshock (see Eq. \eqref{eq:diffOblique}). In order to obtain the appropriate ``mean'' diffusion coefficient for a particle of momentum $p$, we employ an effective $\bm{B^\textrm{eff}_p}(p)$, oriented like the computed $\bm{B_p}(p)$, but with a magnitude \mbox{$|\bm{B_p^\textrm{eff}}(p)|=\max{(|\bm{B_p}(p)|,\delta B_{p}(p))}$}, when making use of the magnetic field amplification due to resonant streaming instability. Here,
\begin{equation}
\delta B_p(p)=\delta B_{1}-\frac{1}{f_1(p)}\int_{-\infty}^0{\textrm{d}x \frac{\textrm{d} \delta B(x)}{\textrm{d}x} f(x,p)} \ ,
\label{eq:deltaBp}
\end{equation}
and $\bm{B_p}$ is given by Eq. \eqref{eq:Bp}. In this way, the gyroradius of a particle in the Monte Carlo simulations is the same as the gyroradius in the expression for the diffusion coefficient (Eq. \eqref{eq:diffOblique}).
Accordingly, we use the appropriate electric fields, as well as mean scattering times. Choosing such a diffusion coefficient appears to be reasonable, considering results from PIC simulations: \cite{caprioli2014c} found that, for strong shocks, the energetic particles experience Bohm diffusion in the magnetic field amplified predominantly by the non-resonant hybrid instability \citep{bell2004}. 

\section{Validation and comparison}
\label{sec:validation}
In this section we compare the results of \cite{caprioli2009} to ours and we further show the results obtained by keeping the parameters of the plasma unchanged, while changing the orientation of the shock. Our treatment differs to some extent from the one in \cite{caprioli2009}. In fact, they account for the possibility that the scattering centres move in the plasma frame with the Alfv{\'e}n speed calculated using the background magnetic field. Nevertheless, they do not find a strong discrepancy between the ``effective compression ratio'' and the MHD compression ratio. For the sake of simplicity, we therefore do not include this effect. We accept solutions where the discrepancy between the pressure calculated with Eq. \eqref{eq:PcAlternative} and with Eq. \eqref{eq:momFluxConsNorm} lies within a $15\%$ tolerance range. Due to the statistical nature of the MC simulations, some fluctuation in the injection efficiency is unavoidable, at every cycle with $r_\textrm{sub}$ fixed, and a tighter tolerance range would require very high statistics, which in turn would require unreasonable computation times. We found that there is a dramatic improvement concerning momentum and energy flux conservation when employing the non-linear approach, as compared to the test-particle set-up. A tighter constraint on the tolerance range would be only a minor correction. Moreover, the uncertainties in the microphysics of acceleration in shocks where the ions reach energies well above their rest mass, as well as the possible development of other instabilities (e.g. the non-resonant hybrid instability) would make it unlikely to improve the reliability of the results by strengthening the convergence criteria. In Table \ref{tab:comparison} we summarize the results of our calculations, with shock obliquities of $0^\circ$, $30^\circ$ and $60^\circ$ and compare to two examples of \cite{caprioli2009}. For all cases, the particle density is $n_0=0.5\times10^6$ m$^{-3}$, the magnitude of the magnetic field is $B_0=5\times 10^{-11}$ T, the temperature is $T_0=10^4$ K, the plasma speed upstream, in the shock frame is $u_0=5.9\times10^6$ m s$^{-1}$, and $\psi=3.7$, as was used in the reference paper.
\begin{deluxetable}{cccccccc}
\tablecaption{Comparison of the solutions of this work, with the semi-analytical results of \cite{caprioli2009}.\label{tab:comparison}}
\tablehead{
\colhead{$\theta_B$} & \colhead{$r_\textrm{sub}$} & \colhead{$r_\textrm{tot}$} & \colhead{$S_\textrm{sub}$} 
& \colhead{$S_\textrm{tot}$} & \colhead{$p_\textrm{max}$}  & \colhead{$T_2$} & \colhead{$RSI$}\\
\colhead{[$^\circ$]} & \colhead{} & \colhead{} & \colhead{} & \colhead{} & \colhead{[$10^6$ GeV/c]} & \colhead{[$10^6$ K]} & \colhead{}
}
\startdata
0 & 3.58 & 112.1 & 3.43 & 108.7 & 0.24 & 0.88 & No \\
0 & 3.84 & 9.22 & 3.79 & 9.12 & 1.17 & 126.5 & Yes \\
\hline
0 & 3.5 & 118 &  &  & 0.25 & 0.78 & No \\
0 & 3.8 & 9.8 &  &  & 1 & 110 & Yes \\
30 & 3.6 & 71 &  &  & 0.25 & 1.4 & No \\
30 & 3.7 & 10.7 &  &  & 1 & 65 & Yes \\
60 & 3.6 & 37 &  &  & 0.25 & 1.6 & No \\
60 & 3.6 & 12.5 &  &  & 1 & 13 & Yes \\
\hline
0 & 2.4 & 17 &  &  & 1 & 5.1 & Yes \\
30 & 2.7 & 17 &  &  & 1 & 7.6 & Yes \\
60 & 2.7 & 17 &  &  & 1 & 2.4 & Yes \\
\enddata
\tablecomments{The semi-analytical results of \cite{caprioli2009} are listed in the first two rows. The authors considered only the parallel configuration. The quantities $S_\textrm{sub}$ and $S_\textrm{tot}$ are the effective compression ratios, when the scattering centres move with the Alfv{\'e}n speed in the plasma frame. The last two rows are the results of the combined approach, using Monte Carlo simulations for the determination of the injection efficiency. All the other rows refer to calculations with $\psi=3.7$. }
\end{deluxetable}
The results for the case of a strictly parallel shock are in good agreement, despite the ``loose'' convergence criteria and the slightly different approach, as mentioned above. 
\ \\

\begin{figure}[ht!]
\tabcolsep15pt\begin{tabular}{cc}
\centering
\includegraphics[width=.43\columnwidth]{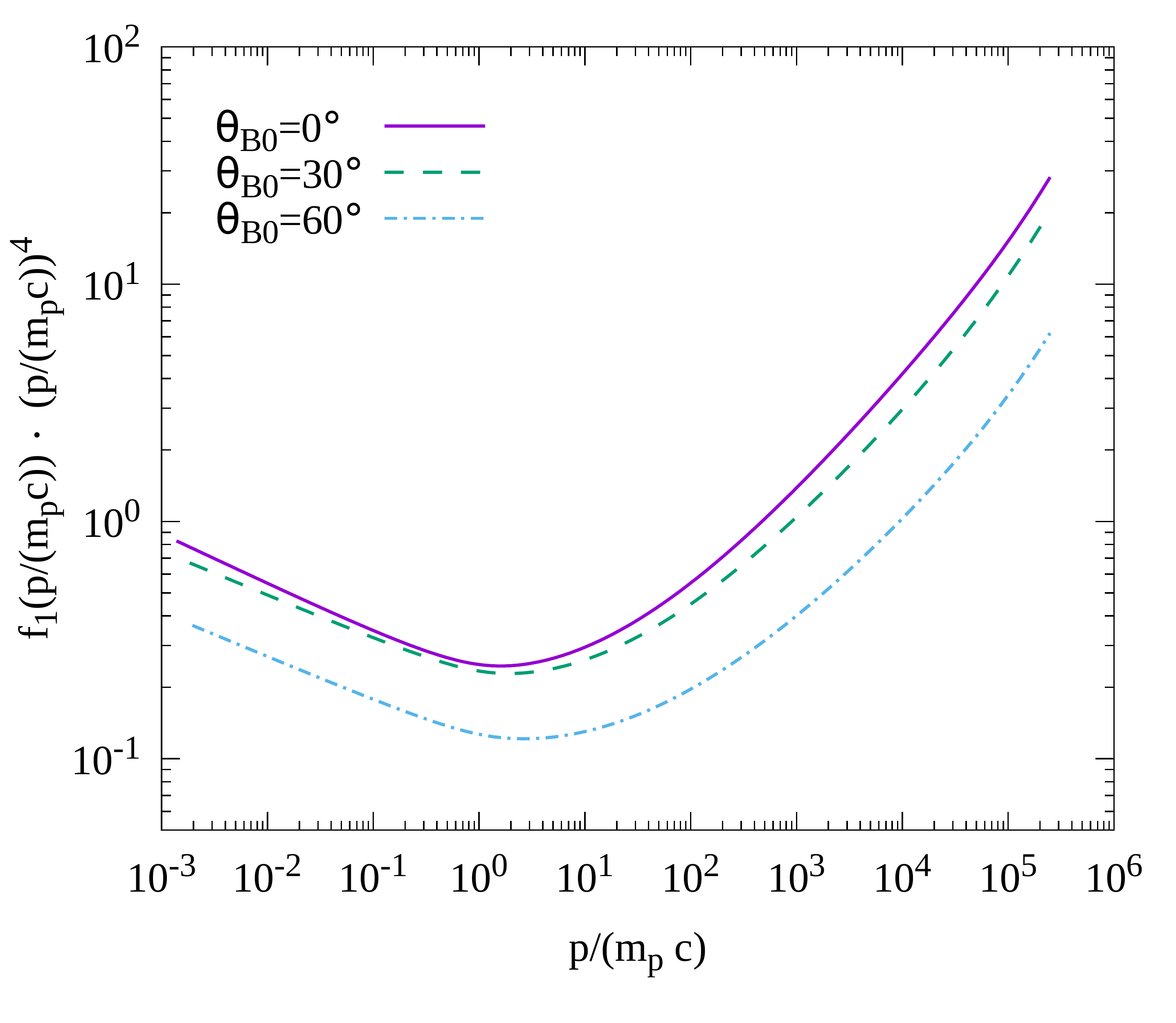}
	& \includegraphics[width=.43\columnwidth]{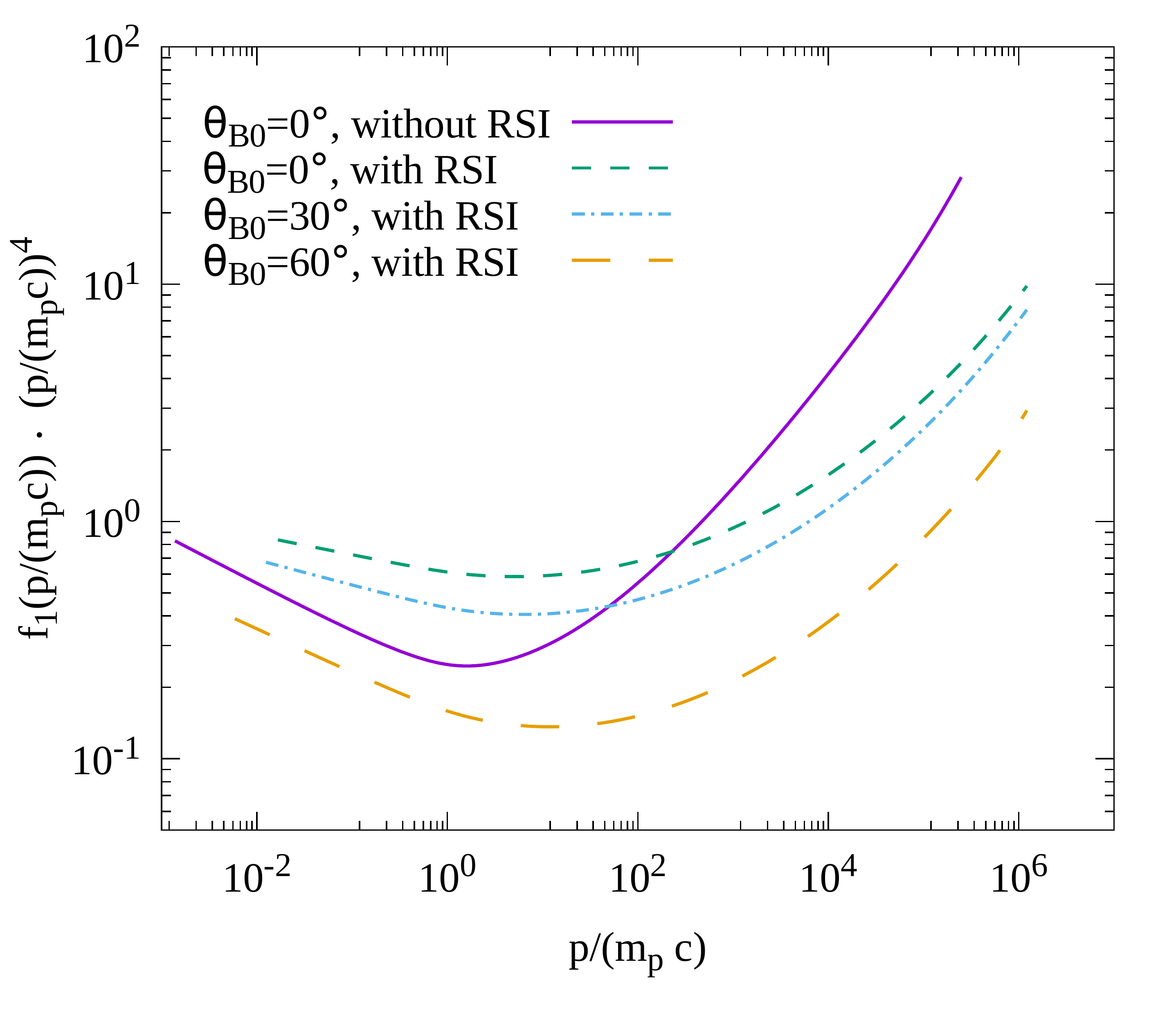}\\
	(a)  & (b)
\end{tabular}
\caption{Non-thermal distribution functions of protons at the position of the subshock of non-linear shocks, multiplied by $[p/(m_p c)]^4$, for three different obliquities. (a) Without magnetic field amplification due to resonant-streaming instability. (b) With the effect of resonant-streaming instability. The result for parallel shock without MFA is plotted as a reference.\label{fig:semianSpectra_ynMFA}}
\end{figure}

In Figure \ref{fig:semianSpectra_ynMFA} (a) we show the distribution function of the non-thermal population of protons obtained by means of the semi-analytical approach alone, for the cases without the RSI effect, summarized in Table \ref{tab:comparison} (rows 3, 5, 7). A trend towards lower densities of accelerated particles is apparent when the shock obliquity increases. This happens despite the (slight) increase of the compression ratio at the subshock, and is an effect of the presence of a non-zero $z$-component of the magnetic field, and the associated pressure: Eqs. \eqref{eq:B(x)} and \eqref{eq:PB} imply that the decrease in the $x$-component of the plasma velocity results in an increase in $B_z$, and in turn in $P_B$. At the subshock, Eq. \eqref{eq:thirdr} must hold, resulting in a lower $r_\textrm{tot}$: the additional magnetic pressure reduces the overall compressibility of the plasma (see Table \ref{tab:comparison}). The downstream temperature is also affected by $P_B$: as expected from Eq. \eqref{eq:T2}, it increases when passing from shock obliquities of $0^\circ$ to $60^\circ$. \ The second plot of Figure \ref{fig:semianSpectra_ynMFA} shows the cases with $P_w\neq0$. Recall that $f_1$ is multiplied by $[p/(m_p c)]^4$. The particle density at the injection momentum $p_\textrm{inj}$ increases with increasing obliquity, due to the increasing compression ratio $r_\textrm{tot}$ and the decreasing downstream temperature. Nevertheless, the decrease of $p_\textrm{inj}$ (caused by the lower $T_2$), combined with the form of the distribution function $f\propto p^{-q}$, is such that the curves at momenta $p\geq p_\textrm{inj,0}$ ($p_\textrm{inj,0}$ being the injection momentum for $\theta_{B0}=0^\circ$) are lower for higher obliquities, resulting in an overall shift of the curves downwards for more oblique cases (Figure \ref{fig:semianSpectra_ynMFA} (b)).

\begin{figure}[ht!]
\tabcolsep15pt\begin{tabular}{cc}
\centering
\includegraphics[width=.43\columnwidth]{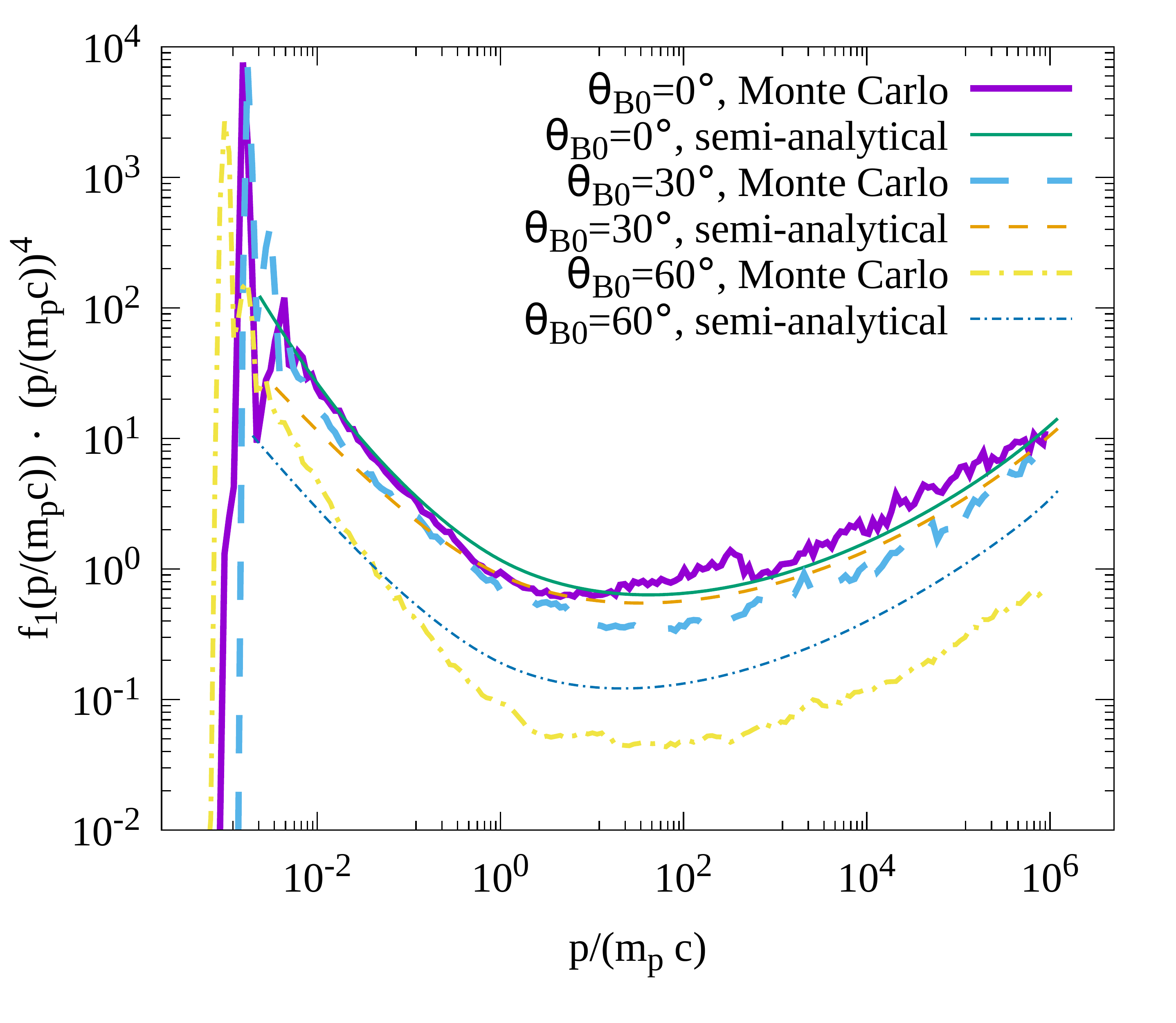}
	& \includegraphics[width=.43\columnwidth]{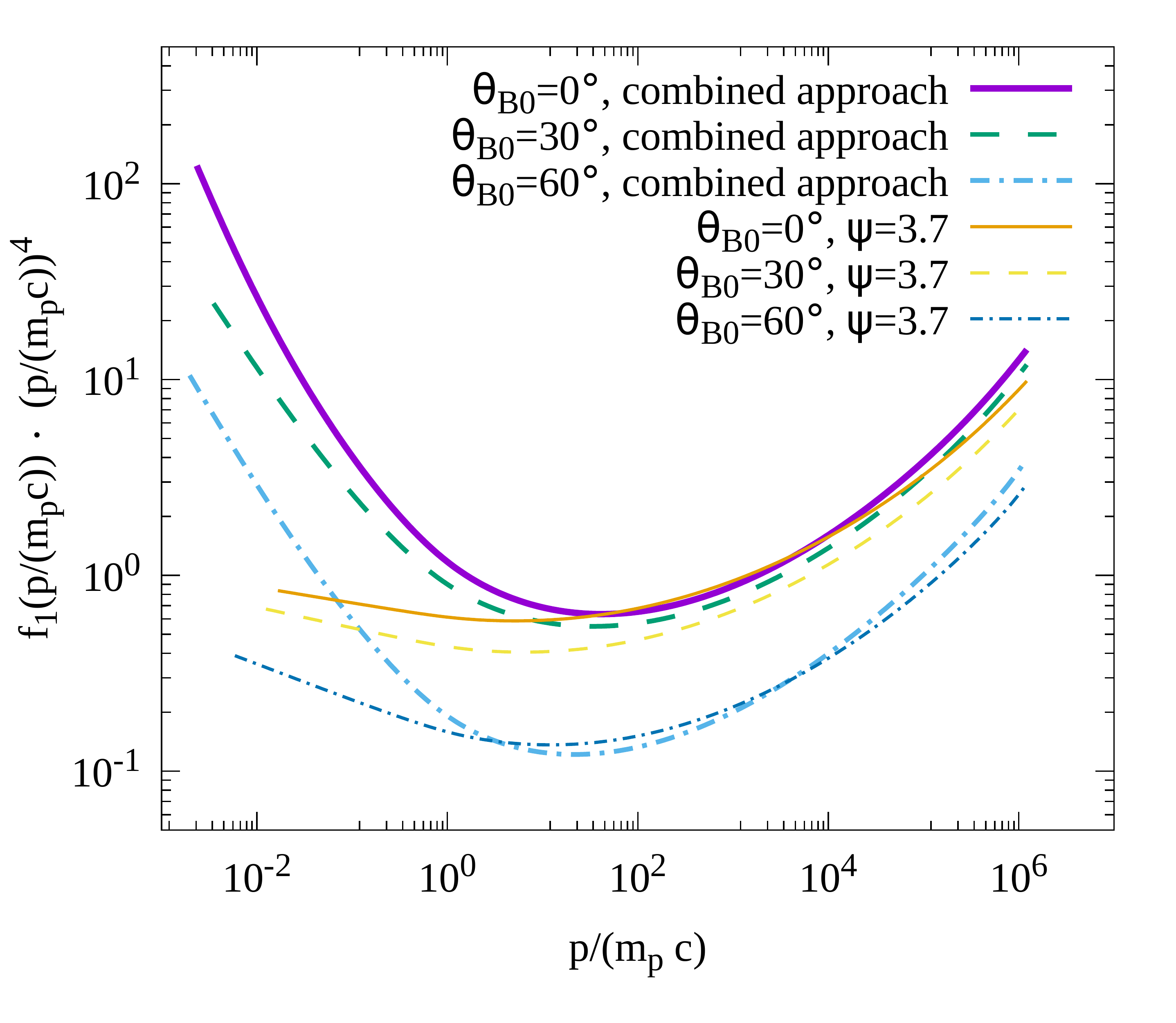} \\
	(a)  & (b)
\end{tabular}
\caption{Non-thermal distribution functions of protons at the position of the subshock of non-linear shocks, multiplied by $[p/(m_p c)]^4$, for three different obliquities. (a) Results from the combined approach employing Monte Carlo simulations and semi-analytical calculations. The curves labelled ``Monte Carlo'' (thick) are the spectra from the full MC simulations after the determination of the non-linear modification of the shock. The curves labelled ``semi-analytical'' (thin) are the solutions obtained with the iterative method described in the text, with Eq. \eqref{eq:f1}.  (b) Comparison between the distribution functions obtained from Eq. \eqref{eq:f1} by employing the combined approach and by fixing $\psi=3.7$.  \label{fig:MCSA}}
\end{figure}

\noindent In Figure \ref{fig:MCSA} (a) we show the results of the approach combining Monte Carlo simulations and semi-analytical calculations of the shock modifications. The curves of the semi-analytical calculations, obtained after determining $\eta$ with the combined approach as described above, and the distributions resulting from the full Monte Carlo simulations with the calculated modified background, are in good agreement. The discrepancy, higher for the oblique shock with $\theta_{B0}=60^\circ$, is lower than a factor of 4. The difference between the spectra of the semi-analytical calculation and the Monte Carlo particle distributions is mainly in the nonrelativistic regime. As already mentioned, a similar feature has also been found in \cite{caprioli2010}. In Figure \ref{fig:MCSA} (b) we compare the semi-analytical non-thermal spectra obtained by fixing $\psi=3.7$ with those employing Monte Carlo simulations for the determination of the injection efficiency. These latter have lower $p_\textrm{inj}$ and much higher injection efficiencies. However, the discrepancies between the two approaches rapidly decrease with increasing momentum, and close to $p_\textrm{max}$ are less than a factor of $\approx 2$. The spectral indices are also different, due to the stronger shock modification in the combined approach (see also Table \ref{tab:comparison}). The similarity at high energies of the solutions with the different injection efficiency determination suggests that the results and discussion presented in Section \ref{sec:results} are quite robust.
\newpage

\section{Results and discussion}
\label{sec:results}
\begin{figure}[ht!]
\tabcolsep15pt\begin{tabular}{cc}
\centering
\includegraphics[width=.43\columnwidth]{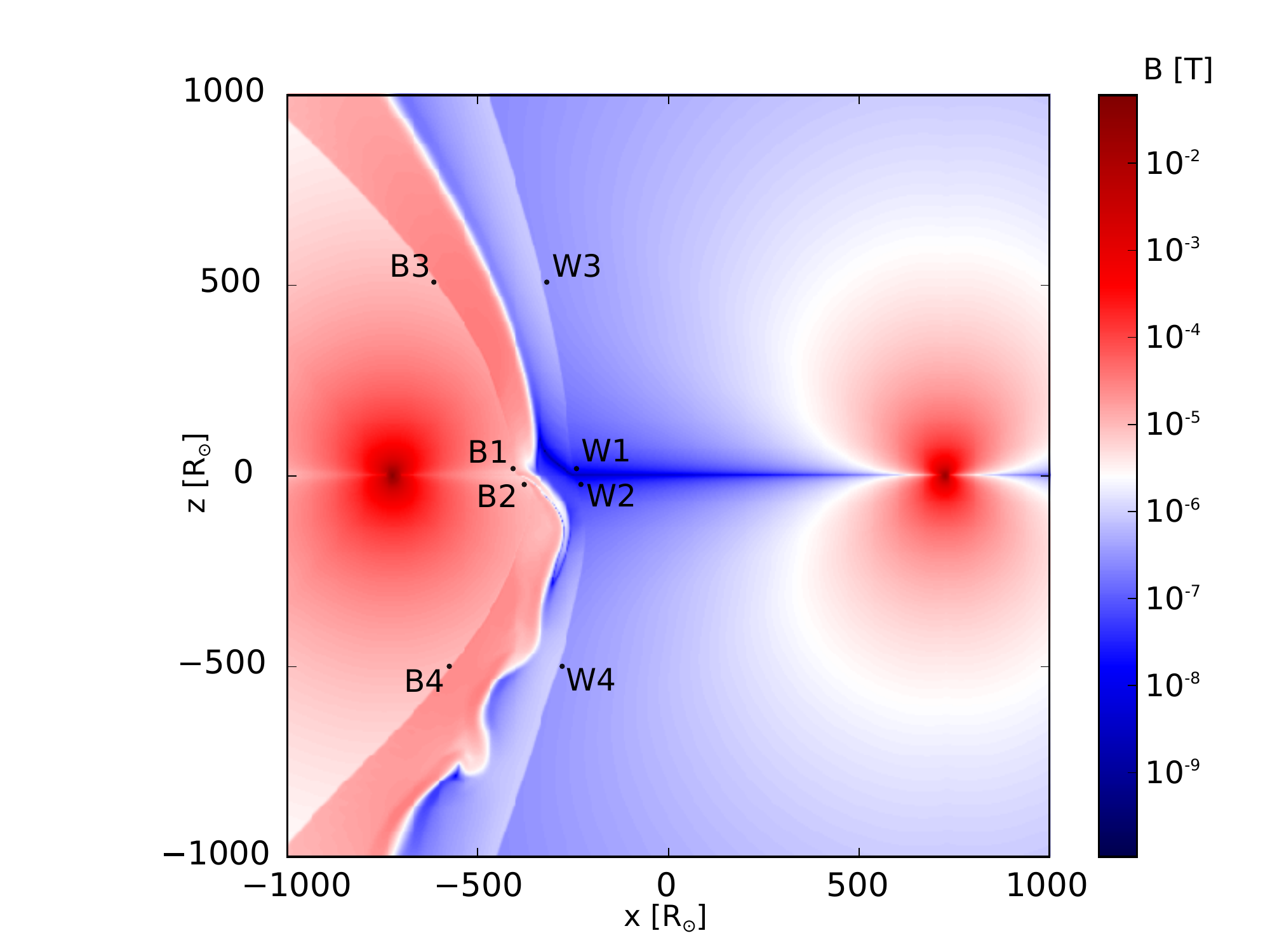}
	& \includegraphics[width=.43\columnwidth]{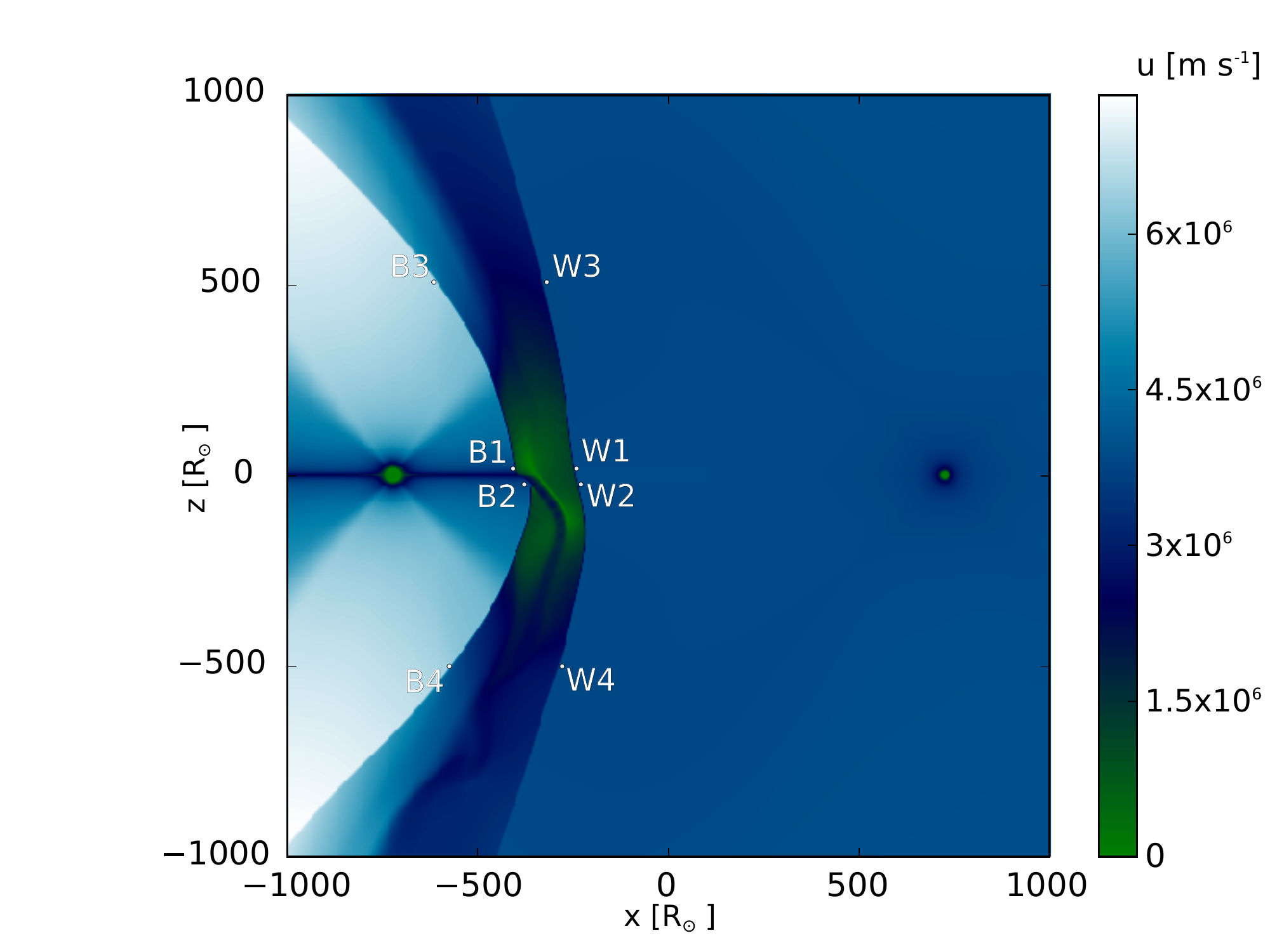} \\
	(a)	&(b)
\end{tabular}
\caption{$x$-$z$ plane of the MHD simulation box and some approximate injection positions (see Table \ref{tab:modifiedShocks}). The B star is on the left, at $x_\textrm{B}=-720 R_\sun$, the WR star is on the right, at $x_\textrm{WR}=720 R_\sun$. (a) Magnetic field strength. (b) Plasma speed.\label{fig:BFieldCWB}}
\end{figure}

In this section, we use the combined method described above to investigate the acceleration of protons in colliding-wind binaries, including a comparison of the test-particle results with the ones that include the back-reaction of the non-thermal protons. The unmodified background is given by a snap-shot of the simulation of an archetypal CWB (see Figure \ref{fig:BFieldCWB}), the parameters of which are given in \mbox{Table \ref{tab:stellarwind}}. The stars in the MHD simulation do not rotate, and there is no orbital motion.
\begin{deluxetable}{cccccccc}[ht!]
\tablecaption{Stellar and wind parameters. \label{tab:stellarwind}}
\tablehead{
\colhead{Star} & \colhead{$M_{*}$} & \colhead{$R_{*}$} & \colhead{$T_{*}$} 
& \colhead{$L_{*}$} & \colhead{$\dot{M}$}  & \colhead{$v_{\infty}$} & \colhead{$B_{*}$}\\
\colhead{} & \colhead{[M$_{\sun}$]} & \colhead{[R$_{\sun}$]} & \colhead{[K]}		& \colhead{[L$_{\sun}$]}		& \colhead{[M$_{\sun}$ yr$^{-1}$]}	& \colhead{[km s$^{-1}$]} 	& \colhead{[G]}
}
\startdata
B		& $30$					& $20$					& $23000$	& $10^5$				& $10^{-6}$						& 4000				& 100\\
WR		& $30$					& $10$					& $40000$	& $2.3\times 10^5$		& $10^{-5}$						& 4000				& 100\\
\enddata
\tablecomments{$M_{*}$ is the stellar mass, $R_{*}$ the stellar radius, $T_{*}$	the effective temperature, $L_{*}$ the luminosity, $\dot{M}$ the mass loss rate, $v_{\infty}$ the terminal velocity of the wind, and $B_{*}$ the surface magnetic field. }
\end{deluxetable}
\noindent The stellar separation is $R=1440 \mbox{ R}_{\sun}$. The region used in the Monte Carlo simulations consists of \mbox{$(x\times y\times z)=(201\times101\times101)$} cubic cells of dimension (3.9$\mbox{ R}_{\sun}$)$^3$. The system is the same as Model A2 of \cite{kissmann2016}.\\
After initializing the background as described in Appendix \ref{sec:app_bg_treatment}, we select 12 super-cells where we inject thermal protons. For the test-particle simulations, we do not modify the background any more and we just let the particles move and scatter in the simulated region. For the simulations including the back-reaction of the accelerated protons, we first find a modified background for the super-cells where the protons are injected, as outlined in Section \ref{sec:methods}. Once a solution is found, we start a simulation embedding the modified super-cells into the same simulated region of the test-particle case. A self-consistent determination of the maximal energies to which the protons are accelerated in the CWB when back-reaction is taken into account would require the modification of the MHD background of the entire WCR, together with the regions upstream of the shocks being modified by the pressure of the non-thermal protons. This is not possible with the approach presented here. \cite{grimaldo2017a} have shown, with test-particle simulations for the same system, that due to the differences in the magnetic field strength on the two sides of the WCR (see Figure \ref{fig:BFieldCWB} (a)), the maximal energies of the accelerated protons can differ by an order of magnitude or more. In fact, on the WR-side, where the magnetic field is weaker, the protons have larger gyroradii and the distribution functions have cutoffs at lower energies ($\approx 10^2\ m_p c$), as compared to the B-side ($\gtrsim10^3\ m_p c$). Therefore, based on results of the test-particle simulations, the maximal momentum of the protons is set to $10^2\ m_p c$ on the WR-side and $10^3\ m_p c$ on the B-side of the WCR. Such an approximation does not take into account that some of the most energetic particles are also accelerated in cells different from the initial one, where the shock is modified. This, combined with the effect described in Section \ref{subsec:semi_anNL}, can lead to an overestimation (more often) or to an underestimation of the proton density at high energies from the semi-analytical calculation, as compared to the final Monte Carlo spectra. However, this does not considerably affect the results of this work.

\begin{figure}[ht!]
\tabcolsep15pt\begin{tabular}{cc}
\centering
\includegraphics[width=.43\columnwidth]{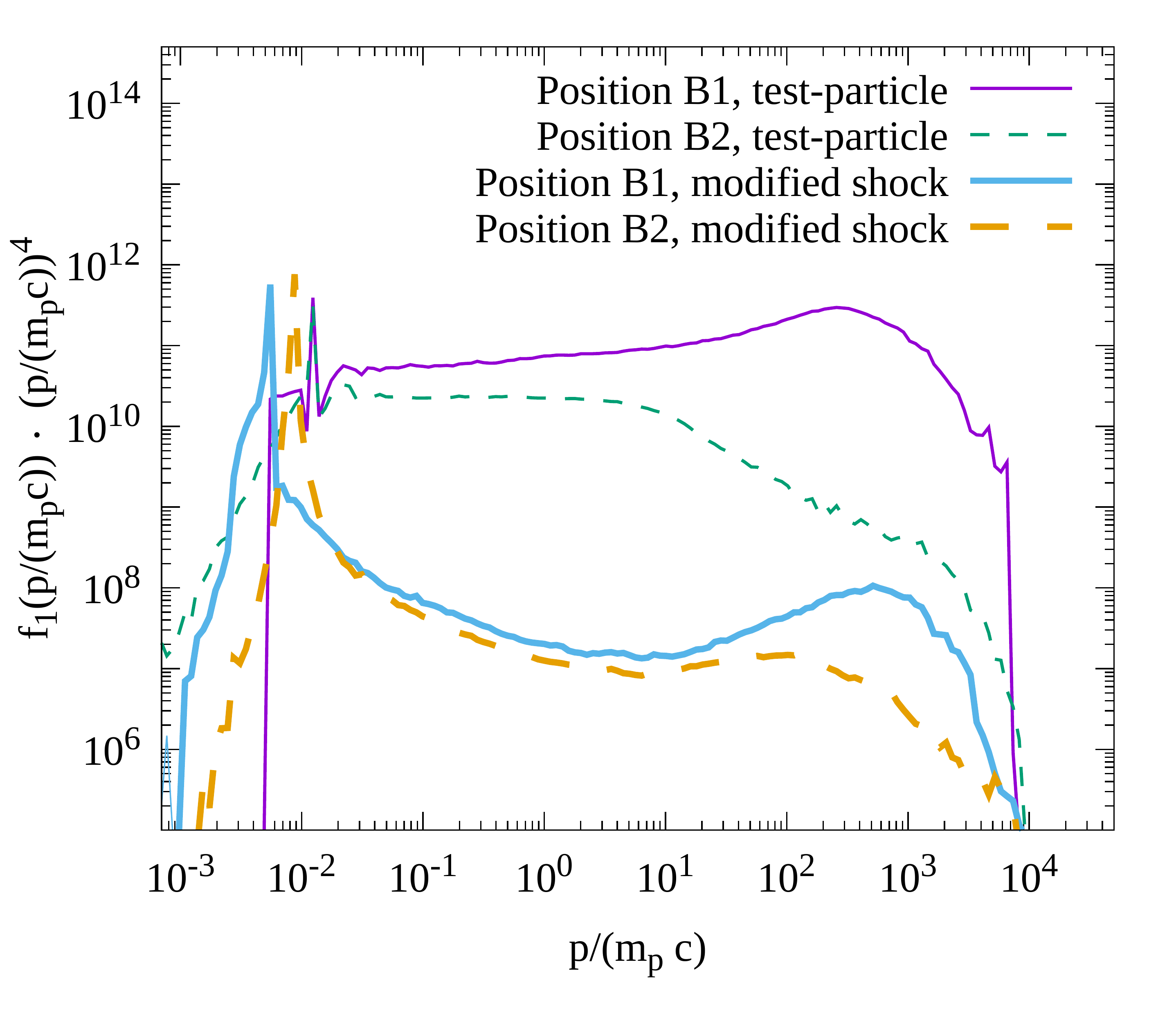}
	& \includegraphics[width=.43\columnwidth]{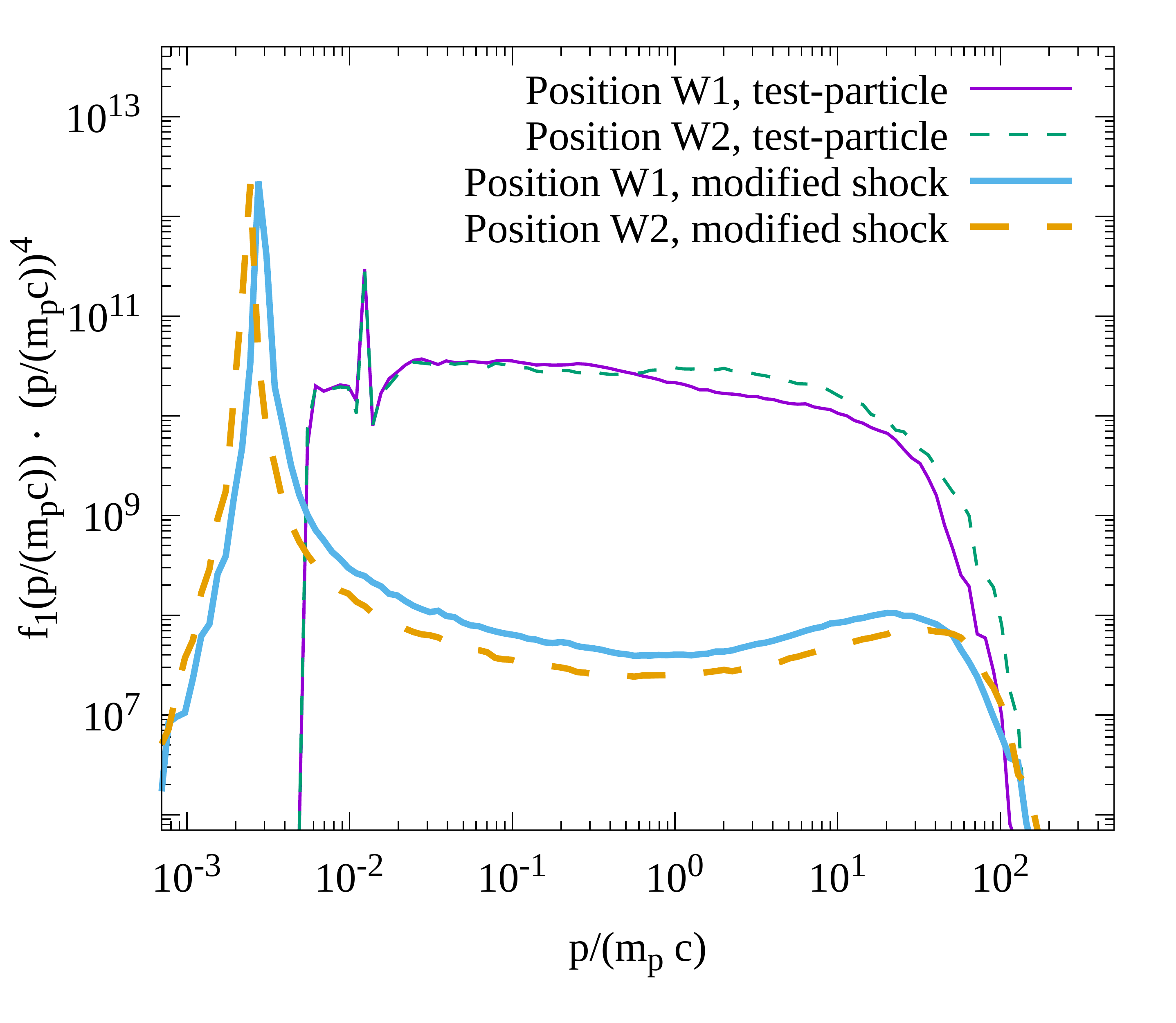}\\
	(a)	&(b)
\end{tabular}
\caption{Proton distribution functions multiplied by $[p/(m_p c)]^4$, resulting from injection of particles at positions close to the apex of the WCR (see Table \ref{tab:modifiedShocks} and Figure \ref{fig:BFieldCWB} for more details), for the test-particle approach (thin lines) and for the non-linear calculations (thick lines). (a) B-side of the WCR. (b) WR-side of the WCR.\label{fig:apexSpectra}}
\end{figure}

\begin{figure}[ht!]
\tabcolsep15pt\begin{tabular}{cc}
\centering
\includegraphics[width=.43\columnwidth]{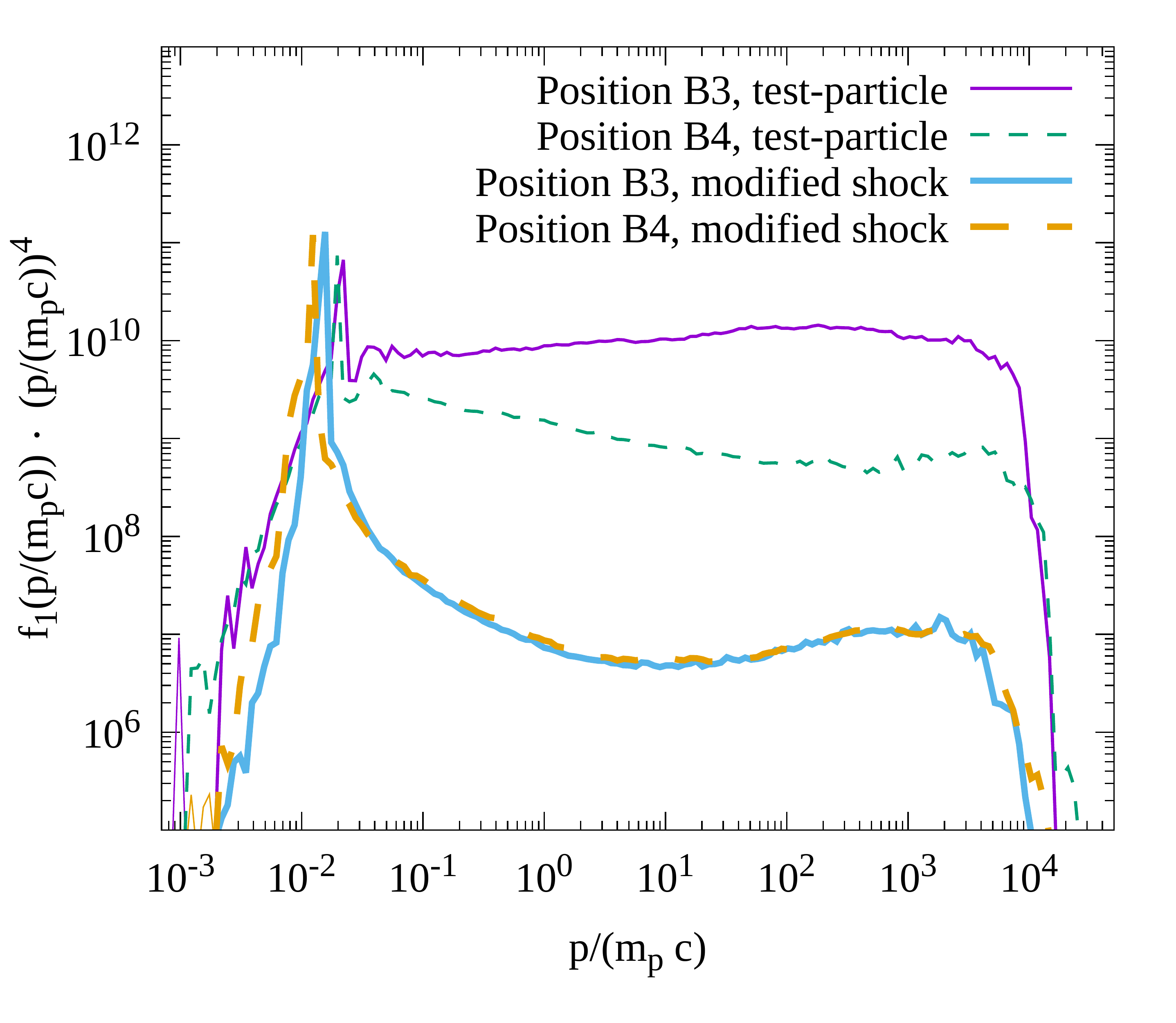}
	&\includegraphics[width=.43\columnwidth]{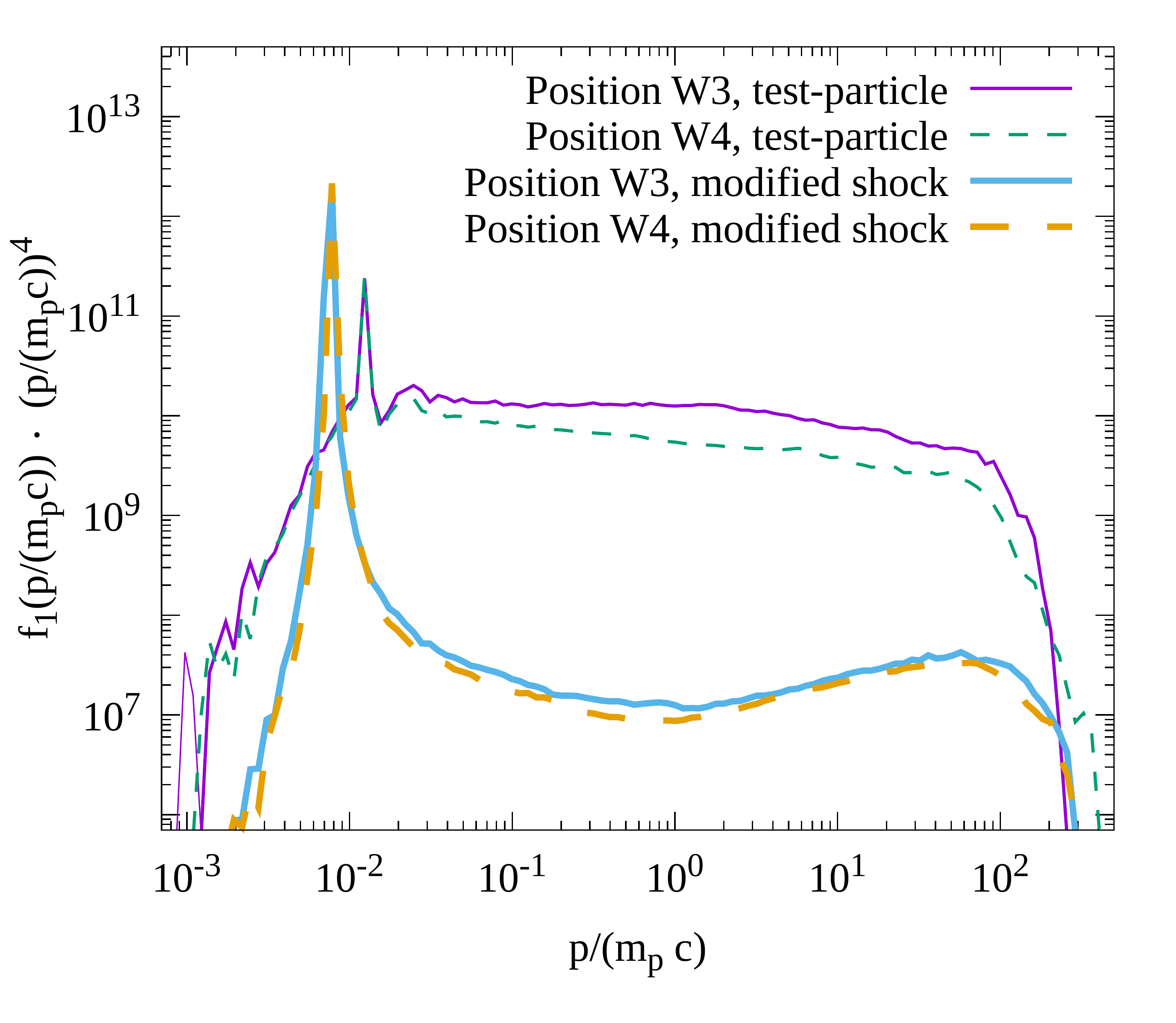} \\
	(a)	&(b)
\end{tabular}
\caption{Same as Figure \ref{fig:apexSpectra}, but at positions farther away from the apex of the WCR (see Table \ref{tab:modifiedShocks} for more details). \label{fig:farSpectra}}
\end{figure}

\begin{figure}[ht!]
\tabcolsep15pt\begin{tabular}{cc}
\centering
\includegraphics[width=.43\columnwidth]{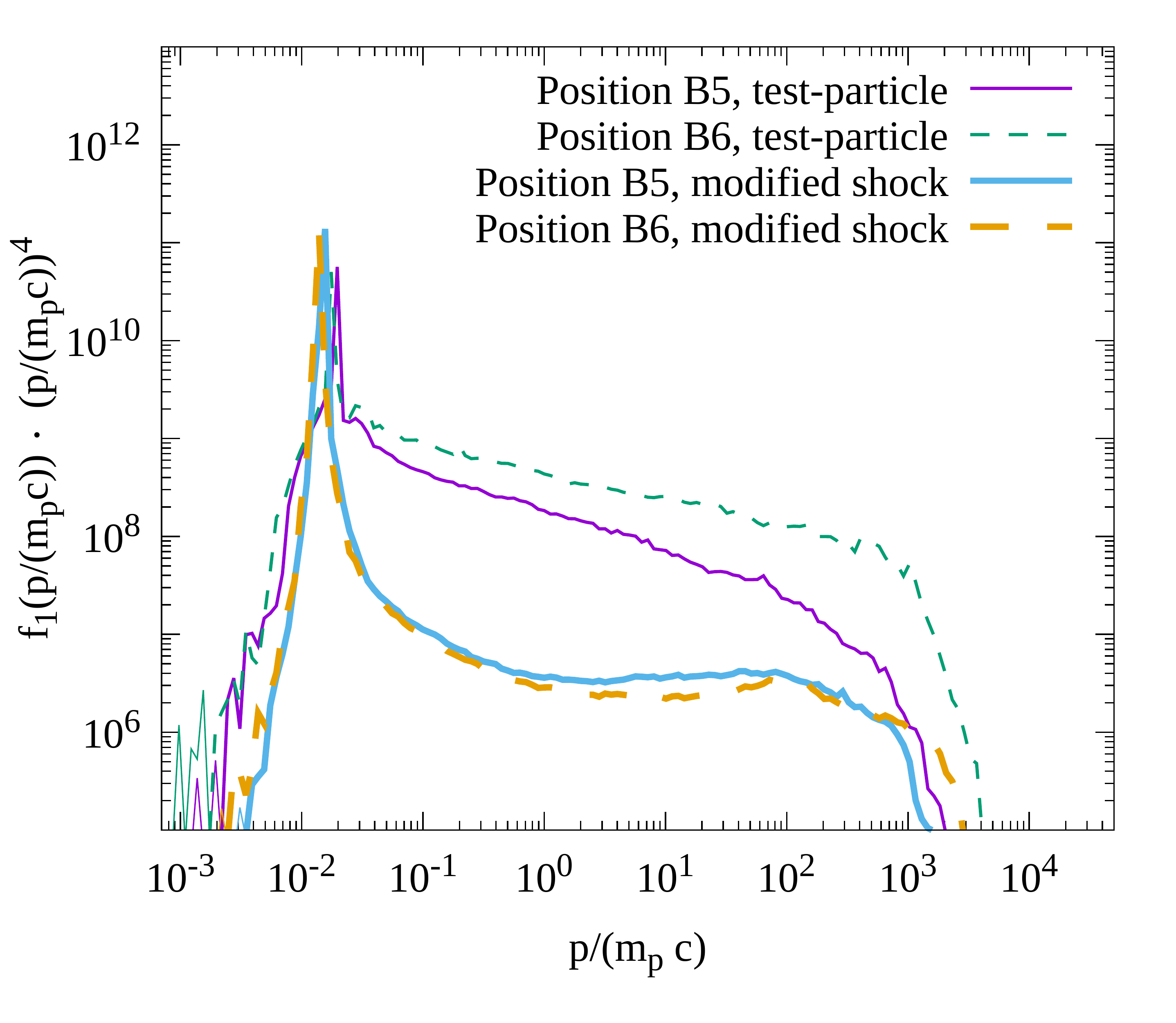}
	&\includegraphics[width=.43\columnwidth]{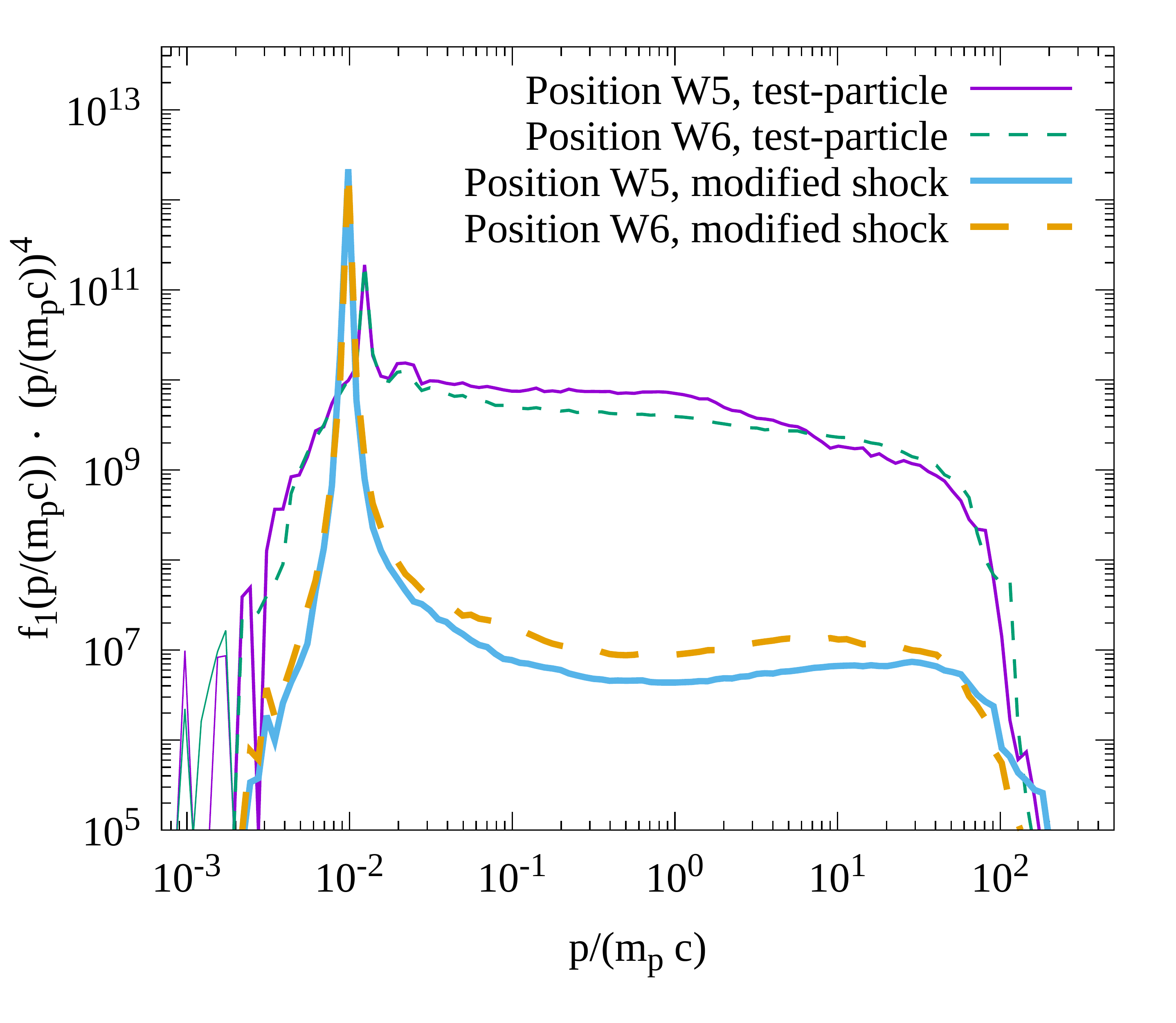} \\
	(a)	&(b)
\end{tabular}
\caption{Same as Figure \ref{fig:apexSpectra}, but at positions farther away from the apex of the WCR (see Table \ref{tab:modifiedShocks} for more details).\label{fig:farSpectra2}}
\end{figure}

\begin{deluxetable}{ccccccc}[ht!]
\tablecaption{Modified shock parameters at different positions along the WCR. \label{tab:modifiedShocks}}
\tablehead{
\colhead{Position} & \colhead{($y,z$)} & \colhead{$r_\textrm{sub}$} & \colhead{$r_\textrm{tot}$} 
& \colhead{$T_2$} & \colhead{$M_{A0x}$} & \colhead{$P_{w1}$} \\
	& \colhead{[$R_\sun$]}	&	&	& \colhead{[$10^6$ K]}
}
\startdata
B1		& (0, 20)	& 2.7					& 9.3	& 10	& 76	& $1.8\times10^{-2}$\\
B2		& (0, -20)	& 2.7					& 9.3	& 6.4		& 62	& $1.7\times10^{-2}$\\
B3		& (0, 420)	& 2.7					& 6.8	& 41	& 30	& $2.2\times10^{-2}$\\
B4		& (0, -420)	& 2.7					& 7.0	& 44	& 33	& $2.3\times10^{-2}$\\
B5		& (500, 420)	& 2.7					& 8.3	& 15	& 43	& $2.9\times10^{-2}$\\
B6		& (-500, -420)	& 2.7  					& 8.4	& 15	& 45	& $2.8\times10^{-2}$\\
W1		& (0, 20)	& 2.6					& 80	& 0.32	& $3.7\times10^{4}$	& $1.1\times10^{-3}$\\
W2		& (0, -20)	& 2.5					& 78	& 0.30	& $3.0\times10^{4}$	& $1.4\times10^{-3}$\\
W3		& (0, 420)	& 2.7					& 33	& 0.80	& $1.7\times10^{3}$	& $6.1\times10^{-3}$\\
W4		& (0, -420)	& 2.6					& 32	& 0.74	& $1.5\times10^{3}$	& $5.2\times10^{-3}$\\
W5		& (500, 420)	& 2.5			& 34	& 0.51	& $1.9\times10^{3}$	& $4.6\times10^{-3}$\\
W6		& (-500, -420)	& 2.7					& 33	& 0.53	& $1.6\times10^{3}$	& $6.9\times10^{-3}$\\
\enddata
\tablecomments{The WR star and the B star are located at coordinates $x_\textrm{WR}=(720, 0, 0)$ and \mbox{$x_\textrm{B}=(-720, 0, 0)$}, respectively. The definitions of the parameters are given in Section \ref{sec:methods}.}
\end{deluxetable}

\begin{figure}[ht!]
\tabcolsep15pt\begin{tabular}{cc}
\centering
\includegraphics[width=.43\columnwidth]{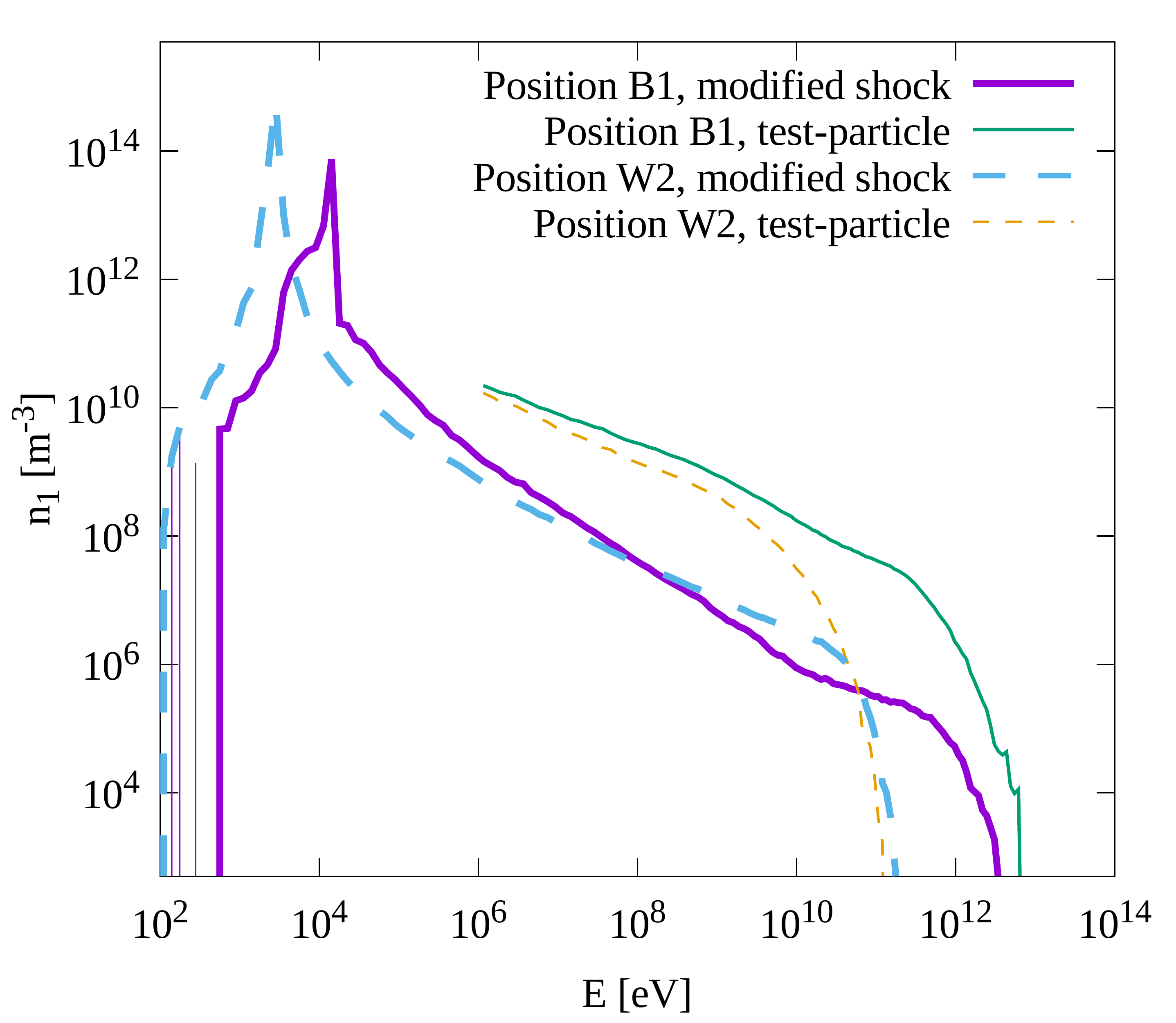}
	&\includegraphics[width=.43\columnwidth]{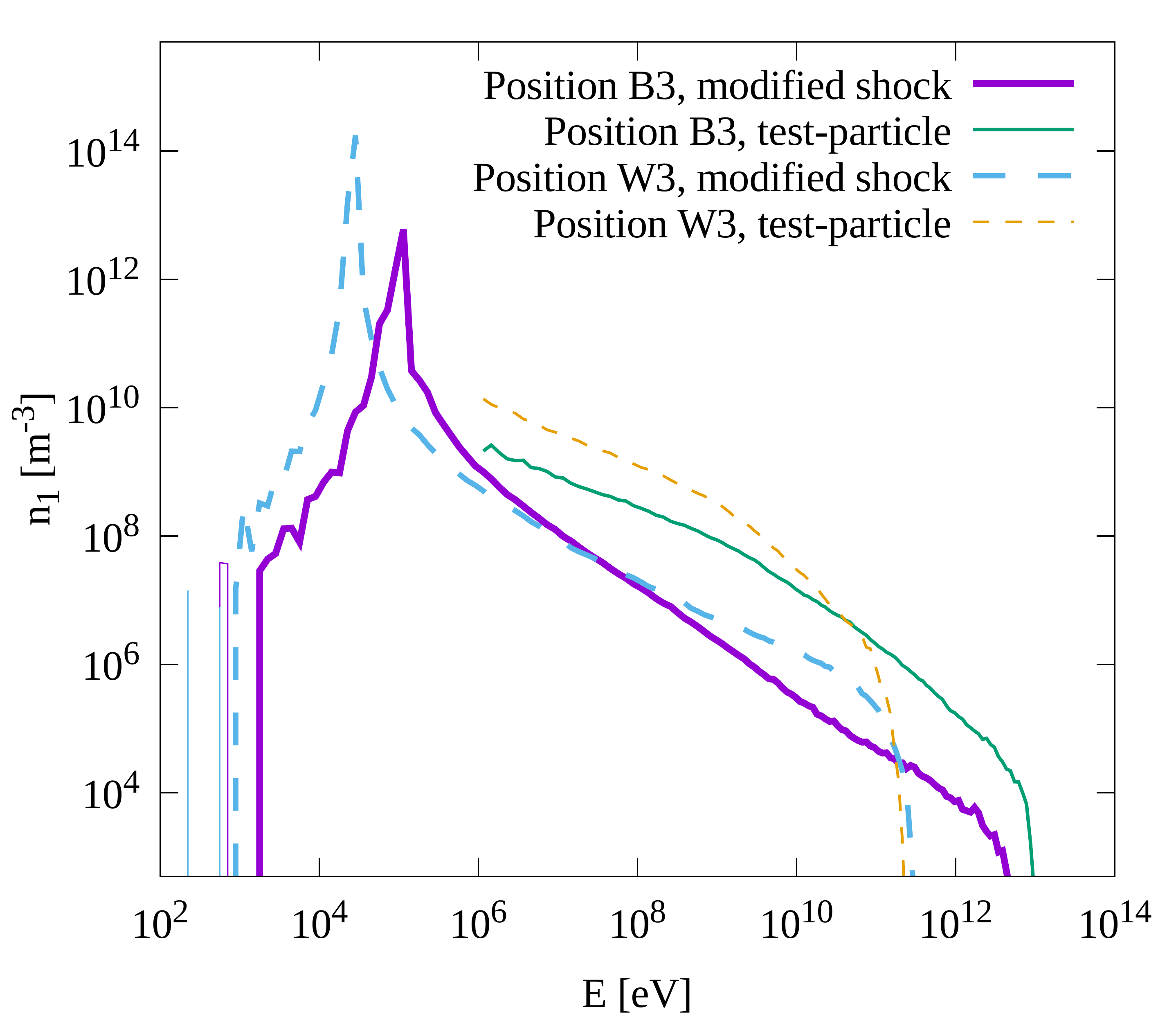} \\
	(a)	&(b)
\end{tabular}
\caption{Proton density spectra as a function of energy obtained by injecting particles at the shocks at four selected positions of the WCR. The thin lines are the test-particle Monte Carlo results, renormalized in order to match the condition $n_1(E=1\textrm{MeV})=10^{-3} n_0$, where $n_0$ is the proton density upstream of the unmodified shock. The thick lines are the non-linear results. (a) Close to the apex of the WCR. (b) Farther away from the apex of the WCR (see Table \ref{tab:modifiedShocks} for more details). \label{fig:renormalized}}
\end{figure}

\noindent In Figure \ref{fig:apexSpectra} (a) we show the distribution functions $f_1(p) [p/(m_pc)]^4$, obtained by injecting the protons close to the apex of the WCR, on the B-side, for both the test-particle and the feedback approach. Note the difference between the test-particle spectra obtained for injection at B1 and at B2. Whereas the spectral indices are very similar at small energies, they become notably different in the relativistic regime, where the B1 spectrum hardens, while the B2 spectrum softens. This is ascribable to different conditions downstream of the respective shocks. The particles which are energetic enough to return to the shock from farther downstream at B1 effectively ``see'' a higher compression ratio, caused by the slowdown of the plasma approaching the contact discontinuity. The effect seen in the test-particle simulations is entirely due to the interaction of the stellar winds, and the geometry of the WCR. In the investigated scenario the downstream flow is slower than it would be for a shock of infinite extent downstream, calculated with the shock-jump conditions. Therefore, it is easier for the particles, to cross the shock again from downstream to upstream in our setup than it would be in a one-dimensional shock structure. We do not see, however, any appreciable effect of hardening of the spectra due to scattering of the particles between the upstream ``colliding shock flows'' on the two sides of the WCR, as modelled by \cite{bykov2013}. This is likely due to different background conditions at, and between, the shocks. \cite{bykov2013} considered a completely symmetric set-up, and obtained a hard spectrum for the distribution function, namely \mbox{$f_1(p)\propto (U_p(p)p)^{-3}$}. For energetic particles with $\lambda_\textrm{mfp}>L_\textrm{WCR}$, where $L_\textrm{WCR}$ is the width of the WCR, it should be possible to see this effect even without considering any modification due to the back-reaction of the accelerated protons upstream of the shocks (i.e. with $U_p(p)=1$). In the system presented here, however, the conditions on the two sides of the contact discontinuity are not equal. Particularly important is the difference in the magnetic field, which is much weaker on the WR-side. As a consequence, the particles accelerated at the B-side shock which manage to cross the contact discontinuity get much larger mean free paths and can therefore more easily escape the system from the WR-side of the WCR. Moreover, the protons accelerated at the WR-side shock do not reach sufficiently high energies that would allow crossing the contact discontinuity and reaching the B-side shock before they are advected out of the simulation box. Therefore, they are not accelerated by scatterings between the two converging flows upstream of the shocks. This situation might change for different parameters of the system, such that the magnetic field strength on both sides of the contact discontinuity is similar. Other factors might also play an important role, e.g. the width and the curvature of the WCR, and the distances between the stars and the shocks. Further studies are required, in order to better understand such effects.\\
As opposed to the B1 spectrum, the B2 spectrum does not harden, but instead it softens in the relativistic regime. This happens because the particles downstream of the shock are more efficiently advected away from the shock once they reach the equatorial plasma flow of the B star which enters the WCR and flows downwards with relatively high velocities (for a discussion concerning the stellar winds and WCR structure in this and other systems, see \cite{kissmann2016}).  \\
When comparing the test-particle spectra with those obtained with shock modification shown in Figure \ref{fig:apexSpectra} (a), the effect of the back-reaction of the accelerated protons is clearly visible. The density of the non-thermal protons is reduced by up to more than three orders of magnitudes. The shock modification is stronger on the WR-side, as can be seen in Figure \ref{fig:apexSpectra} (b) and in Table \ref{tab:modifiedShocks}: the ``thermal peak'' moves towards lower momenta, while the total compression ratio reaches much higher values on the WR-side. This is caused by the very high Alfv{\'e}n Mach number on the WR-side. The Alfv{\'e}n waves produced via the resonant streaming instability lower considerably the value of $r_\textrm{tot}$, but not as much as on the B-side of the WCR, where the pressure associated to the waves is up to about one order of magnitude larger (see Table \ref{tab:modifiedShocks}). In Figures \ref{fig:farSpectra} and \ref{fig:farSpectra2} we show the same as in Figure \ref{fig:apexSpectra}, but for different positions. In all cases the test-particle results strongly overestimate the acceleration efficiency, as was found in many studies of SNRs (e.g. \cite{vladimirov2009}, \cite{bykov2014}, \cite{amato2005}, \cite{kang2012}).\\
A self-consistent quantitative modelling of the $\gamma$-ray emission from real systems is beyond the scope of this paper. Instead we conjecture the impact of back-reaction on the estimated $\gamma$-ray fluxes from CWB systems. The lack of detection of $\gamma$-rays from CWBs is well known. For example, \cite{werner2013} did not find any evidence for $\gamma$-ray emission from various systems (amongst which, WR 11, WR 140, WR 147 have been considered), despite the predictions of several models (e.g. \cite{eichler1993,benaglia2003,reimer2006}). Besides $\eta$ Carinae, $\gamma^2$ Velorum (WR 11) has been recently detected as a $\gamma$-ray source. In the latter, the dominant $\gamma$-ray production channel is probably the decay of neutral pions produced in nucleon-nucleon collisions, because of synchrotron and inverse Compton losses in the magnetic and radiation fields from the stars at the WCR \citep{reitberger2017}. In the model of \cite{reimer2006}, the dominant production mechanism for $\gamma$-rays in WR 140 and WR 147 was found to be inverse Compton scattering from relativistic electrons in the stellar radiation field. This picture might considerably change if particle acceleration is efficient and magnetic field amplification takes place, leading to a situation similar to what has been found by \cite{reitberger2017}. There, the inverse Compton losses in the strong radiation fields prevent the electrons from reaching sufficiently high energies for $\gamma$-ray production. The $\gamma$-ray emission is therefore likely of hadronic origin. In the same work it is further claimed that a stronger magnetic field and smaller inverse Compton losses would also yield $\gamma$-ray fluxes in agreement with observations.\\
As far as the normalization of the non-thermal tail of the particle distribution is concerned, an injection parameter is commonly used. In \cite{reitberger2017}, for example, the proton density at 1 MeV is set to $n(E=1\textrm{MeV})=\eta_{1\textrm{MeV}} n_0$, with the injection parameter $\eta_{1\textrm{MeV}}=10^{-3}$, and $n_0$ the proton density of the background plasma. In Figure \ref{fig:renormalized} we show the particle density for a selection of injection positions obtained with feedback, together with the non-thermal tails of the test-particle case, shifted in order to match the injection parameter of \cite{reitberger2017}. Amongst B1-B6 and W1-W6, we chose the positions corresponding to the higher densities at high energies in the test-particle case, close to the apex and far away from it, for each side of the WCR. We see that the actual particle densities might be even lower than what was obtained with $\eta_{1\textrm{MeV}}=10^{-3}$. 
Together with the spectral energy distributions of the particles, the $\gamma$-ray fluxes from a modified WCR are also expected to change, but not necessarily in the same manner: the increased density in the WCR (higher compression ratios) will also have an effect, since hadronic collisions producing neutral pions will be more frequent there. In the following, we roughly estimate the $\gamma$-ray production for the model studied here, in the local vicinity of the apex of the WCR, using as the target proton population the thermal particles downstream of the considered shock. At position B1, the ratio between the non-thermal proton density (Eq. \eqref{eq:defn1}) in the model with the modified shock, $n_1^\textrm{NL}(60\textrm{ GeV})$, and the one in the test-particle approach, with $\eta_{1\textrm{MeV}}=10^{-3}$, $n_1^\textrm{TP}(60\textrm{ GeV})$, is  $n_1^\textrm{NL}(60\textrm{ GeV})/ n_1^\textrm{TP}(60\textrm{ GeV}) \approx 0.02$. At an energy of $10$ GeV (corresponding to a proton energy of $E_p\approx E_\gamma / \kappa_{\pi^0} \approx 60$ GeV, with the inelasticity factor for pion production $\kappa_{\pi^0}\simeq 0.17$ \citep{aharonian2000}), the ratio between the $\gamma$-ray emissivities would be $q_\gamma^\textrm{NL}(10\textrm{ GeV})/q_\gamma^\textrm{TP}(10\textrm{ GeV})\approx0.07$. At the WR-side of the WCR, at position W2, the ratios of the densities of the non-thermal population at $\approx 6$ GeV is $n_1^\textrm{NL}(6\textrm{ GeV})/ n_1^\textrm{TP}(6\textrm{ GeV}) \approx 0.06$, while the $\gamma$-ray emissivity in the case of a modified shock, due to the much larger compression ratio, would be higher than in the test-particle case: $q_\gamma^\textrm{NL}(1\textrm{ GeV})/q_\gamma^\textrm{TP}(1\textrm{ GeV})\approx2$. Despite being a limiting case, since the non-linear compression ratio close to the apex on the WR-side reaches the highest values, this estimate highlights the non trivial modifications of the $\gamma$-ray fluxes due to non-linear modifications of the shocks of the WCR in colliding-wind binary systems.\\
Another intriguing observable effect of proton acceleration in CWBs, even if highly speculative, might be the change of the opening angle of the WCR-cone. \cite{reitberger2017} found that modelling $\gamma^2$ Velorum with a ``strong coupling'' between the stellar winds, i.e. including the radiative braking of the wind of the \mbox{O star} due to the photon field of the WR star, yields an opening half-angle of the shock-cone ($\approx72^\circ$), i.e. closer to the observed value of $\approx 85^\circ$, as compared to the case when radiative braking is neglected ($\approx24^\circ$). High-energy protons escaping the WCR from the O-side towards the WR-side, might slow down even further the wind of the WR star, and therefore contribute to a further shock-cone opening. Indeed, in our simulation we observe that some of the protons with higher energies eventually reach the contact discontinuity at the interface between the B-side and the WR-side downstream regions. There, the magnetic field changes abruptly, being weaker on the WR-side, increasing the mean free path of the particles and enhancing their escape probability on that side. \\
The discussion above shows that a viable way to reconcile the theoretical predictions for $\gamma$-ray fluxes with the observations of CWBs in the radio and $\gamma$-ray bands, is to employ the widely accepted idea of the existence of a back-reaction of the accelerated particles at collisionless shocks.

\section{Conclusions and outlook}
\label{sec:conclusions}

In this work we have presented a combined approach, employing magnetohydrodynamic simulations, a semi-analytical method for obtaining non-linear solutions of modified collisionless shocks, and Monte Carlo simulations of proton acceleration. In order to use the methods usually applied to strictly parallel shocks to the broad variety of obliquities found along the wind-collision regions of colliding-wind binaries, we adapted the equations to the case where the shock normal and the magnetic field are not aligned. By applying our method to a model of a typical CWB, we showed that, similarly to what was found in studies of typical SNR shocks, the Monte Carlo test-particle results differ considerably from the non-linear solutions: the test-particle approach greatly overestimates the injection efficiencies, which are dramatically reduced when energy and momentum conservation at the shock is fulfilled. Remarkably, we found indications that the injection and acceleration efficiencies at the shocks of CWBs may be lower than what is often assumed in approaches based on the transport equation for the accelerated particles, neglecting their back-reaction on the shocks. In the test-particle approximation, the maximal energies reached by the protons are different, mainly depending on which side of the WCR they were injected. On the B-side, energies of up to almost $10$ TeV can be reached, while on the WR-side the cut-off is in the range of \mbox{$10$-$100$ GeV}. This is ascribed to the different strengths of the magnetic fields. We note, however, that this difference might be reduced if the shock modifications were globally taken into account, due to an stronger magnetic field amplification on the WR-side, which reduces the difference of the magnetic field strength. The total compression ratios differ systematically from the B-side to the WR-side, the latter being much higher, due to a smaller magnetic turbulence pressure at the subshocks on that side. Also based on the results of \cite{reitberger2017}, we formulated the hypothesis that magnetic field amplification due to the accelerated protons could increase synchrotron losses of electrons accelerated at the shocks delimiting the WCR. This would reduce the maximal energy reached by the relativistic electrons, preventing them from efficiently producing $\gamma$-rays via inverse Compton scattering in the stellar photon fields. This might help to explain why non-thermal synchrotron emission has been observed by many CWB systems, while so far there has been no detection of $\gamma$-rays from those sources, with only one exception so far, that is $\gamma^2$ Velorum ($\eta$ Carinae does not show any synchrotron emission in the radio domain, presumably due to synchrotron self-absorption.). Further and deeper investigations of this and other observable effects of non-linear shock modifications in colliding-wind binaries, including the application to observed CWB systems, will be subject of future studies.

\acknowledgments

E. G. and A. R. acknowledge financial support from the Austrian Science Fund (FWF), project \mbox{P 24926-N27.} E.G. further acknowledges the receipt of a ``Doktoratsstipendium aus der Nachwuchsf\"orderung'' (2017/3/MIP-3) from the University of Innsbruck.

\appendix

\section{Background treatment} \label{sec:app_bg_treatment}

\begin{figure}[h]
\centering
\tabcolsep15pt\begin{tabular}{cc}
\centering
\includegraphics[width=.43\columnwidth]{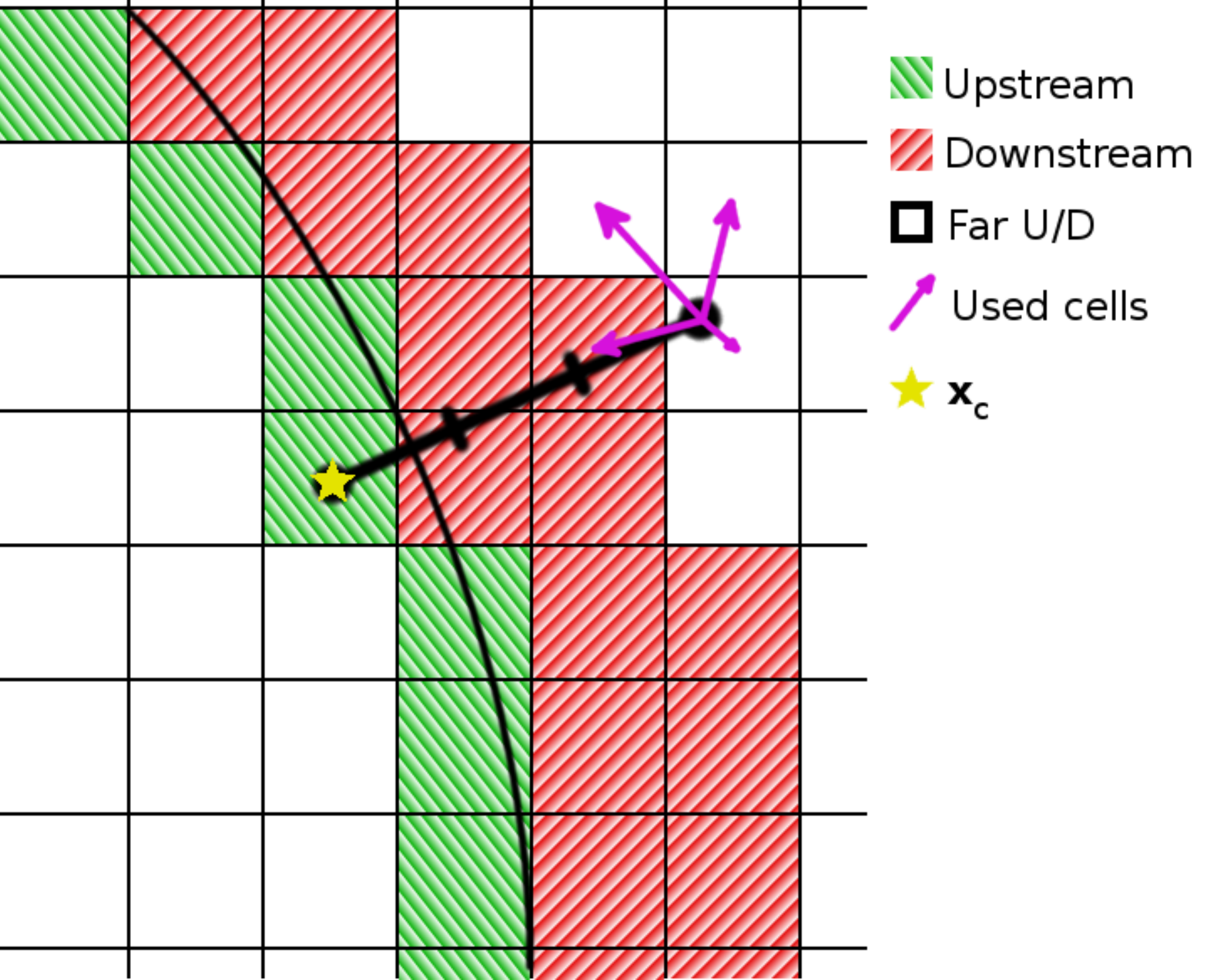}
  & \includegraphics[width=.43\columnwidth]{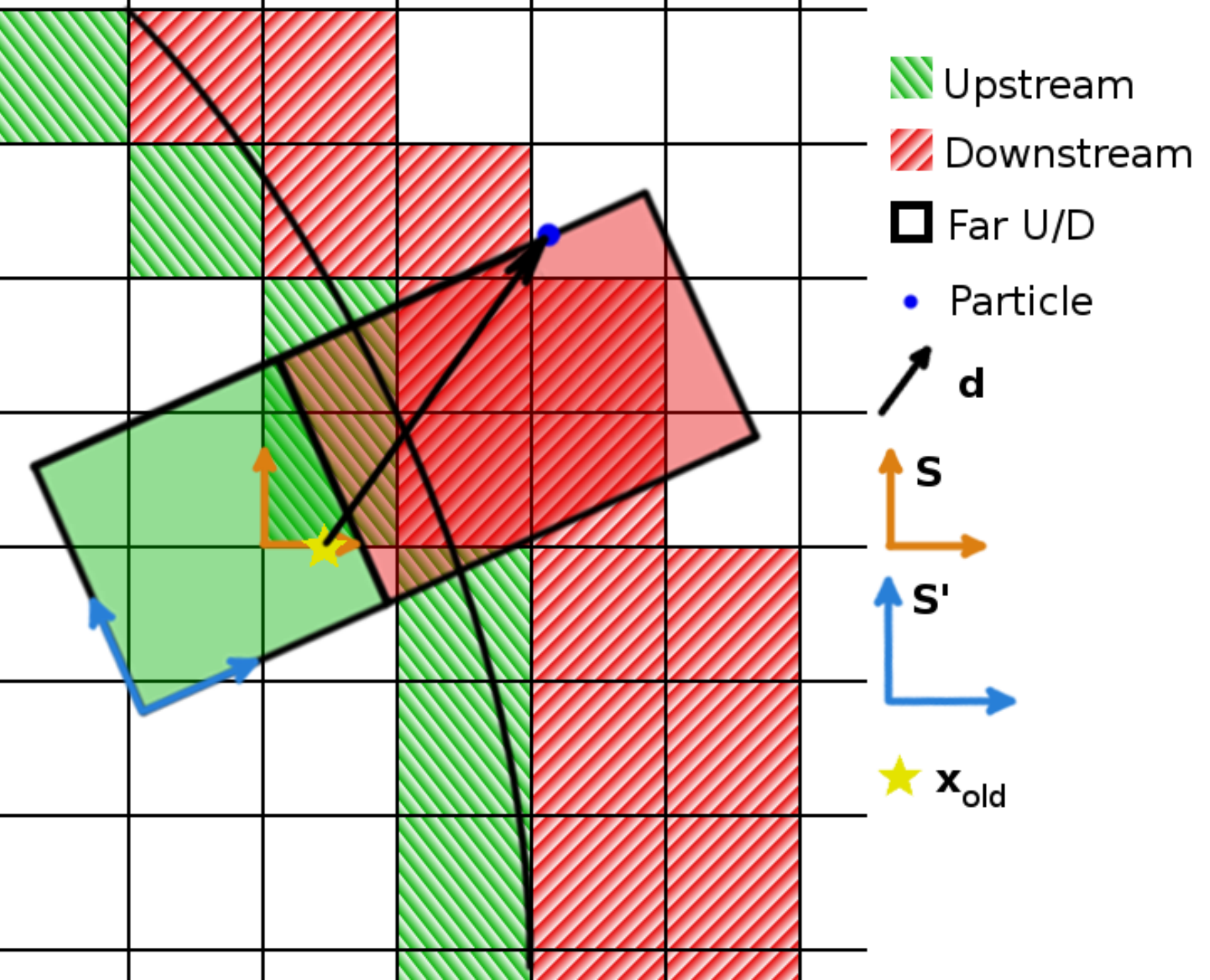}  \\
  (a) \qquad \qquad \qquad \qquad \qquad & (b) \qquad \qquad
\end{tabular}
\label{fig:bg}
\caption{(a) Illustration of the method used for the initialization of the background of the downstream super-cells. These latter result from a weighted average of the fields within a distance $\Delta x$ from the point 3$\Delta x$ downstream of the centre $x_c$ of the associated upstream shock-front cell, in the direction normal to the shock front ($\Delta x$ is the cell size). (b) Schematic representation of a pair of upstream and downstream super-cells. When a particle enters a cell marked as ``shock front'' cell, its position $x_{\textrm{old}}$ in the simulation domain is recorded, so that, when leaving the super-cell regime, the new position in the normal background is found by adding the displacement vector $\vec{d}$ to the recorded coordinates. $\bf{S}$ is the non-rotated reference frame, $\bf{S'}$ is the rotated reference frame used in the super-cell regime. Cell sizes are exaggerated for display purposes; super-cell sizes are $(2\Delta x)^3$ upstream, and $3\Delta x \times 2\Delta x \times 2\Delta x $ downstream.}
\end{figure}

Here we will give details of the approach applied in order to use the MHD results as the background for the particle acceleration simulations. We justify first the need for a modification of the ``raw'' outputs.\\
In the MHD modelling, the transition between upstream and downstream of a shock is about three cells wide. When simulating CWBs, this layer is much bigger than the mean free path of the thermal particles. As a consequence, the compression ratio ``seen'' by the protons injected into the MHD background would be much different from the actual one, which would in turn effectively prevent particles from being accelerated. The Monte Carlo simulations run under the assumption that the scattering centres of the thermal particles are comoving with the background plasma, with a discontinuity at the position of the shock (or the subshock, in the case of non-linear modifications). Therefore, we chose to set up a system of superimposed cells, of size $(2\Delta x)^3$ upstream, and $3\Delta x \times 2\Delta x \times 2\Delta x $ downstream, with a sharp jump between the fields of the two cells ($\Delta x$ is the size of a cell in the MHD simulation). In Figure \ref{fig:bg} (a), we depict the procedure for the test-particle approach, described in the following.\\
First of all, we identify the position of the shock as mentioned in Section \ref{subsec:MC}, as well as its orientation. We then mark as ``upstream cells'' those being immediately before the WCR, in the direction pointed by the plasma flow, and as ``downstream cells'' the first two cells following the shock. We associate each upstream cell to a couple of upstream-downstream super-cells. The background upstream is determined by averaging the fields of the respective MHD upstream cell and the neighbouring upstream-marked ones. The background downstream is a weighted average of the fields of the cells within a radius $\Delta x$ from the point $3\Delta x$ downstream of the upstream MHD cell \citep{reitberger2014a} . The super-cell fields are then rotated so that the shock surface an the cell boundary between the upstream super-cell and the downstream one have the same orientation. This is necessary, since that boundary \textit{is}, for the simulated particles, the shock surface. In the non-linear case, we initialize the super-cell couple where the protons are injected with the fields obtained as described in Section \ref{subsec:semi_anNL}.\\
During the simulation, when a particle enters a shock front cell (upstream or downstream), its position is saved and the simulation is performed with a two-cells setup, until the particle leaves the super-cells. At this point, the position of the particle in the whole simulation box is calculated, starting from the previously saved position and adding the (appropriately rotated) total displacement in the super-cells. The simulation is then continued in the ``normal'' MHD background until the end or until it enters again the shock front domain.

\section{Wave pressure} \label{sec:app_wavepressure}

Here, we summarize the derivation of the pressure associated with the Alfv{\'e}n waves in the case of an oblique shock. Following \cite{caprioli2009}, we start from the stationary equation for growth and transport of magnetic turbulence which, for an oblique shock, reads:
\begin{equation}
\frac{\partial \mathcal{F}_w(k,x)}{\partial x} = u_x(x) \frac{\partial \mathcal{P}_w(k,x)}{\partial x} + \sigma(k,x)\mathcal{P}_w(k,x)\ .
\label{eq:turbulence}
\end{equation}
Herein, $\mathcal{F}_w$ is the energy flux, $\mathcal{P}_w$ is the pressure, both per unit logarithmic bandwidth, $\sigma$ is the growth rate of the energy in the magnetic turbulence.
The latter is given by \citep{skilling1975c}, and for the case of only backwards propagating waves, as assumed here and in many other works (e.g. \cite{mckenzie1982},  \cite{kang2007}, \cite{caprioli2012}) it reads:\\
\begin{equation}
\sigma(k,x) = \frac{\pi^2 m_p^2 \Omega_0^2 v_A}{B_0^2} 2\pi \iint d\mu\ dp\ p^2 (1-\mu^2) v^2 \frac{\bm {\hat n_B}\cdot\bm\nabla f(x,p)}{\nu_-} \delta(kp|\mu|-m_p\Omega_0)\ , 
\label{eq:growthRate}
\end{equation}
where $\Omega_0 = qB_0/m_p$ is the nonrelativistic gyrofrequency, $B_0$ is the (mean) background magnetic field, $\bm {\hat n_B}$ is the unit vector along $B_0$, $\nu_- = \pi\Omega_0\mathcal{P}_{w}/(4\gamma U_B)$ is the collision frequency of particles against waves moving forward (backwards) in the frame comoving with the plasma. In the latter expression, $\gamma$ is the Lorentz factor of the particle, while $U_B=B_0^2/(2\mu_0)$ is the magnetic energy density of the background field. 
Integrating Eq. \eqref{eq:turbulence} over $k$, and normalizing by $\rho_0 u_{0x}^2$, yields:
\begin{equation}
2U_x(x)\frac{\textrm{d} P_w(x)}{\textrm{d}x} = V_{Ax}(x) \frac{\textrm{d} P_c (x)}{\textrm{d}x} - 3 P_w(x) \frac{\textrm{d} U_x(x)}{\textrm{d}x}\ ,
\label{eq:diffPw}
\end{equation}
where $V_{Ax}(x)=v_A(x)\cos\theta_B(x)/u_{0x}$. As discussed in \cite{caprioli2009}, we can neglect $P_B$, $P_w$ and $P_g$ in Eq. \eqref{eq:momFluxConsNorm} in the precursor (but not at the subshock) with respect to the kinetic momentum flux and the pressure of the accelerated particles, if acceleration is efficient, and use $P_c(x)\simeq 1-U_x(x)$. The solution of Eq. \eqref{eq:diffPw}, assuming no wave pressure far upstream is:
\begin{equation}
P_w(x)=U_x(x)^{-\frac{3}{2}}\left [ \frac{(1-U_x^2(x))\ \cos\theta_{B0}}{4 M_{A0x}} \right ] \ .
\label{eq:alpha(x)_app}
\end{equation}

\bibliographystyle{apj}%
\bibliography{references}%

\end{document}